\documentclass[twocolumn,showpacs,preprintnumbers,amsmath,amssymb]{revtex4}
\usepackage{epsfig}
\usepackage[utf8]{inputenc}

\usepackage[breaklinks=true]{hyperref}

\usepackage[table]{xcolor}
\usepackage{graphicx}
\usepackage{dcolumn}
\usepackage{bm}
\usepackage{esvect}

\usepackage{hyperref}
\hypersetup{colorlinks=true,linkcolor=magenta,anchorcolor=green,citecolor=cyan,filecolor=black,menucolor=black,urlcolor=brown}
\usepackage{slashed}

\newcommand{\be}{\begin{equation}}
\newcommand{\ee}{\end{equation}}
\newcommand{\ba}{\begin{eqnarray}}
\newcommand{\ea}{\end{eqnarray}}

\begin{document}

\title{Subsonic accretion and dynamical friction for a black hole moving through a self-interacting scalar dark matter cloud}

\author{Alexis Boudon\footnote{\href{alexis.boudon@ipht.fr}{alexis.boudon@ipht.fr}}}
\affiliation{Universit\'{e} Paris-Saclay, CNRS, CEA, Institut de physique th\'{e}orique, 91191, Gif-sur-Yvette, France}
\author{Philippe Brax}
\affiliation{Universit\'{e} Paris-Saclay, CNRS, CEA, Institut de physique th\'{e}orique, 91191, Gif-sur-Yvette, France}
\author{Patrick Valageas}
\affiliation{Universit\'{e} Paris-Saclay, CNRS, CEA, Institut de physique th\'{e}orique, 91191, Gif-sur-Yvette, France}

\begin{abstract}

We investigate the flow around a black hole moving through a cloud of self-interacting scalar
dark matter.
We focus on the large scalar mass limit, with quartic self-interactions, and on the subsonic regime.
We show how the scalar field behaves as a perfect gas of adiabatic index $\gamma_{\rm ad}=2$
at large radii while the accretion rate is governed by the relativistic regime close to the
Schwarzschild radius. We obtain analytical results thanks to large-radius expansions,
which are also related to the small-scale relativistic accretion rate.
We find that the accretion rate is greater than for collisionless particles, by a factor $c/c_s \gg 1$,
but smaller than for a perfect gas, by a factor $c_s/c \ll 1$, where $c_s$ is the speed of sound.
The dynamical friction is smaller than for a perfect gas, by the same factor $c_s/c \ll 1$,
and also smaller than Chandrasekhar's result for collisionless particles, by a factor
$c_s/(cC)$, where $C$ is the Coulomb logarithm. It is also smaller than for fuzzy dark matter,
by a factor $v_0/c \ll 1$.

\end{abstract}

\date{\today}

 \maketitle

\section{Introduction}
\label{sec:introduction}

The standard model of cosmology, $\Lambda$CDM, has two main components  whose nature is still undetermined today:  dark energy - $\Lambda$ - and  Cold Dark Matter - CDM. Dark matter is commonly described by a nonrelativistic perfect fluid present both on galactic and cosmological scales. It is necessary for the formation and evolution of structures as observed today, from
galactic to cosmological scales. The most recent observations predict an energy fraction  of $\approx 0.25$ for dark matter \cite{Planck:2018vyg, DES:2022qpf}, about 5 times more than for ordinary matter.   Weakly Interacting Massive Particles (WIMPs), which could originate from the electroweak sector of physics beyond the standard model, are still one of the strongest contenders despite the absence of relevant signal from particle physics and direct/indirect detection experiments.

Notwithstanding its  great success on large astrophysical and cosmological scales, the nature of dark matter is still mysterious. Moreover, there are now tensions between the $\Lambda$CDM paradigm  and  observations, amongst which the most celebrated are the  core-cusp problem, the missing satellites and the too-big-to-fail problems \cite{DelPopolo:2016emo, Hui:2001wy, deBlok:2009sp, Weinberg:2013aya}. Although these discrepancies can be  partially resolved by taking into account baryonic feedback,  as confirmed by numerical simulations, the jury is still out and  this may not be enough to resolve completely  the current tensions \cite{Dutton:2018nop, Dutton:2020vne}. Similarly,  the latest tests of the WIMP hypothesis, as provided by direct detection, are getting closer and closer to the neutrino floor and the possibility of a direct experimental confirmation of the existence of WIMPs seems more and more remote \cite{Liu:2017drf, Billard:2021uyg}.

Alternative scenarios have been proposed, e.g. scalar-field dark matter (SFDM), where dark matter consists of spin-0 bosons or bound fermions, whose masses $m$ are within the range $10^{-22} $ eV to $1$ eV \cite{Hu:2000ke,Hui:2016ltb,Battaglieri:2017aum, Ferreira:2020fam, Khoury:2021tvy}. One striking feature of these models is the existence of static and stable configurations comprising a large number of scalar degrees of freedom and amenable to a semi-classical description \cite{Grobov:2015lda,Hui:2016ltb,Khoury:2021tvy}.
These solitons lead to dark matter halos with flat cores that could address the core-cusp problem. On the other hand, even if these solitons are not relevant for the galactic-scale tensions,
e.g. if $m \gg 10^{-22} \, {\rm eV}$ and their radius is smaller than $1 \,\rm kpc$,
they could still have an impact on the dynamics of other orbiting objects in the halo.

For instance, when an astrophysical object interacts with a large collection of other compacts bodies, its motion is slowed down by dynamical friction. The collisionless case was first treated by Chandrasekhar \cite{Chandrasekhar:1943ys}. This phenomenon also happens when a compact objects such as a black hole (BH) penetrates inside a dark matter soliton, or more generally a fluid medium. The case of a fuzzy dark matter (FDM) halo has been analysed in \cite{Hui:2016ltb} and corresponds to dark matter models with no self-interactions. In this paper, we will consider the case of a scalar dark matter model where self-interactions are important. The self-interactions play two prominent roles. First of all, the solitons are modified as compared to the FDM case, as
equilibrium configurations correspond to a balance of gravity by the scalar pressure associated with
the self-interactions, whereas for FDM gravity is balanced by the so-called quantum pressure at the scale of the de Broglie wavelength. Second, the nature of the dynamical friction changes drastically as we find out in this paper.
However, the dynamical friction experienced by the BH is also different from the case
of a perfect gas, because the scalar field has a specific behavior in the nonlinear and
relativistic regime close to the BH horizon.

In this paper, we concentrate on the dynamical friction of a Schwarzschild BH moving inside
a dark matter soliton, as this could be relevant to current research on gravitational waves \cite{Gomez:2017wsw, DeRosa:2019myq}.
The calculation of the dynamical friction in FDM systems has already been thoroughly studied \cite{Hui:2016ltb, Berezhiani:2019pzd, Hartman:2020fbg,Wang:2021udl}.
In the low-velocity regime that we consider in this paper, the dynamical friction is  greater than the one  for
collisionless particles  \citep{Chandrasekhar:1943ys} or for a perfect gas
\cite{Ostriker:1998fa,Lee:2011px} (but it is smaller in the more usual high-velocity regime).
We will show in this paper that for self-interacting SFDM the dynamical friction is greatly
reduced.
The dynamical friction induces a dephasing on the emission frequency of gravitational waves studied for Extreme and Intermediate Mass Ratio Inspirals (EMRIs and IMRIs), which may be detected by future detectors such as eLISA and DECIGO \cite{Macedo:2013qea, Barausse:2014tra, Cardoso:2019rou, Li:2021pxf}. The difference of dephasing  between the FDM case and SFDM is left for future work and could serve as a distinguishing feature of the SFDM scenario which  may  be detectable experimentally. In another context, the Fornax globular clusters timing problem, i.e. the observed tension in the orbital decay of clusters of stars as seen in the Fornax Dwarf Spheroidal, could be addressed by SFDM models. Indeed, CDM numerical simulations lead to a faster orbital decay, and so a higher dynamical friction, than obtained in observations (e.g., \cite{Bar:2021jff, S_nchez_Salcedo_2022} as reviews). For FDM, this tension disappears for low DM masses $m<10^{-21}$ eV \cite{Safarzadeh:2019sre, Lancaster:2019mde, Hartman:2020fbg}, but this range of scalar masses is in potential tension with other observables. In the SFDM, dynamical friction is also expected to be lower than in CDM and this could help in resolving this tension.

In this study, we focus on the subsonic case where the velocity of the incoming BH does not exceed the speed of sound in the scalar halo.
Within the bulk of the scalar cloud, where at equilibrium the gravitational potential
$\Phi_{\rm N}$ is balanced by the effective pressure $P$ associated with the self-interaction,
the circular velocity of compact objects inside the gravitational well is
$v_c^2 \sim \Phi_{\rm N} \sim P/\rho \sim c_s^2$.
Therefore, the BH typically moves at a relative velocity of the order of $c_s$.
Thus, the subsonic regime $v_0 < c_s$ covers both the low-velocity limit (e.g., when the BH
has dissipated its kinetic energy and fallen into the center of the gravitational potential)
and the lower half of the typical cases $v_0 \sim c_s$. As we expect no singularity at the
threshold $c_s$, it should also provide the correct order of magnitude for all cases
$v_0 \sim c_s$.
In addition to its practical relevance, the subsonic case also provides a relatively simpler
context to lay out the problem, the main equations of motion and the method of solution,
in comparison with CDM and FDM.
The case of supersonic motion, where shocks appear and make the analysis much more
intricate, will be treated in a forthcoming paper.

We find that there is a strong connection between the dynamical friction and the accretion rate of dark matter into the BH, i.e. the rate of infalling matter into the BH. In fact, the dynamical friction is both proportional to the accretion rate and to the relative velocity. Such a linear velocity dependence is familiar as it also occurs at low velocity for dynamical friction in a gas cloud with a non-vanishing speed of sound \cite{Lee:2011px}.
In our case too, the SFDM solitons are characterized by a non-vanishing speed of sound
\cite{Brax:2019fzb}.
For larger velocities in the relativistic regime, several studies already exist when self-interactions are not present or subdominant \cite{Hartman:2020fbg,Traykova:2021dua,Vicente:2022ivh}. When interactions are present, the case of a gas with a non-vanishing speed sound will only be the lowest order approximation as the self-interactions make the dynamics nonlinear and new effects could be expected, especially in the relativistic regime. This is left for future work.

The outline of this paper is as follows.
In Sec.~\ref{sec:dark-matter} we introduce scalar-field dark matter with a quartic self-interaction.
We introduce the action, the equations of motion in the nonrelativistic limit, and their equilibrium
solutions (that we denote solitons). These will also set the boundary conditions far from the BH.
We briefly discuss the comparison with FDM and we obtain the parameter space where
our derivations apply.
In Sec.~\ref{sec:relativistic_regime}, we present the fully relativistic equation of motion, which is
required to obtain the boundary condition near the Schwarzschild radius and to determine
the global solution. We explain how we derive the solution in the large-mass limit, fully taking into
account the nonlinearity due to the self-interaction. We recall that in the radial case
\cite{Brax:2019npi} there are two velocity branches and that, as for the perfect fluid, the boundary
conditions select the unique transonic solution and its critical flux.
We compare our derivation with those for FDM, which rely on the linearity of the Klein-Gordon
equation when there are no self-interactions.
In Sec.~\ref{sec:eq_of_motion_phase}, we focus on radii much greater than the Schwarzschild radius and show how the equations of motion match those of an isentropic fluid.
In Sec.\ref{sec:flow_SFDM_BH}, we obtain the flow of SFDM around the BH in the subsonic regime,
using an iterative numerical scheme. We also introduce large-distance expansions.
These allow us to obtain explicit expressions for the accretion rate onto the BH in
Sec.~\ref{sec:accretion}, and for its dynamical friction in Sec.~\ref{sec:friction}.
We conclude in Sec.~\ref{sec:Conclusion}.
Some complementary derivations are given in the appendices.

\section{Dark matter scalar field}
\label{sec:dark-matter}

\subsection{Scalar-field action}
\label{sec:scalar-action}

In this paper, we study the scalar-field dark matter model governed by the following  action
\be
S_\phi = \int d^4x \sqrt{-g} \left[ - \frac{1}{2} g^{\mu\nu} \partial_\mu\phi
\partial_\nu\phi - V(\phi) \right] ,
\label{eq:Action}
\ee
where we include a quartic self-interaction,
\be
V(\phi) = \frac{m^2}{2} \phi^2 +  V_{\rm I}(\phi) \;\;\; \mbox{with} \;\;\;
V_{\rm I}(\phi) = \frac{\lambda_4}{4} \phi^4 , \;\;\; \lambda_4 > 0 .
\label{eq:V_I-def}
\ee
The coupling constant $\lambda_4$ is taken positive to ensure that the self-interaction is repulsive (a negative sign corresponds to attractive self-interaction). This leads to an effective
pressure that can counter-balance gravity and lead to static and stable dark matter
halos on small scales, called solitons in the following.

On the cosmological background or on galactic scales, the oscillations of the scalar field due to the
quadratic mass term are required to be dominant,
leading to  an upper bound on $\lambda_4$.
This ensures that, at lowest order, the scalar field behaves as cold dark matter with a vanishing
pressure. Then, the interaction term is a small perturbation that slightly modifies the harmonic
oscillations of the scalar field and gives rises to an effective pressure, which leads to deviations
from the CDM scenario on small scales.
In particular, this leads to a characteristic scale \citep{Brax:2019fzb}
\be
r_a = \sqrt{\frac{3\lambda_4}{2}} \frac{M_{\rm Pl}}{m^2} ,
\label{eq:ra-def}
\ee
where $M_{\rm Pl}$ is the reduced Planck mass.
This sets both the Jeans length, which is independent of density and redshift
\citep{Goodman:2000tg}\cite{Chavanis:2011uv} and below which density perturbations
of the cosmological background cease to grow and oscillate, and the size of hydrostatic
equilibria (solitons) that can form after collapse and decoupling from the Hubble expansion.
In the nonrelativistic regime, which applies to large scales in the late Universe and to astrophysical
scales far from BH horizons, one can decompose the solutions to the nonlinear Klein-Gordon
equation between the fast oscillations at frequency $m$ and a slowly varying envelope that evolves
on cosmological or astrophysical timescales. The latter is then governed by the Schr\"odinger
equation.
We refer the reader to \cite{Brax:2019fzb} for a cosmological study of these SFDM scenarios.
In the following, we focus on subgalactic scales and discard the expansion of the universe.

\subsection{Nonrelativistic regime}

In the nonrelativistic weak-gravity regime, it is convenient to write the real scalar field
$\phi$ in terms of a complex field $\psi$ as
\be
\phi = \frac{1}{\sqrt{2 m}} \left( e^{-i m t} \psi + e^{i m t} \psi^\star \right) .
\label{eq:phi-psi}
\ee
In this regime, where typical frequencies $\dot\psi/\psi$ and momenta $\nabla\psi/\psi$
are much smaller than $m$, the complex scalar field $\psi$ obeys the Schr\"odinger equation,
\be
i \, \dot\psi = - \frac{\nabla^2\psi}{2m} + m ( \Phi_{\rm N} + \Phi_{\rm I} ) \psi ,
\label{eq:Schrodinger}
\ee
where $\Phi_{\rm N}$ is the Newtonian gravitational potential and $\Phi_{\rm I}$ is the
nonrelativistic self-interaction potential. For the quartic self-interaction it reads \cite{Brax:2019fzb}
\be
\Phi_{\rm I} = \frac{m |\psi|^2}{\rho_a} \;\;\; \mbox{with} \;\;\; \rho_a =
\frac{4 m^4}{3\lambda_4} .
\label{eq:Phi_I-psi}
\ee
It is also convenient to express $\psi$ in terms of the amplitude $\rho$ and the phase $s$
by the Madelung transform \cite{Madelung_1927},
\be
\psi = \sqrt{\frac{\rho}{m}} e^{i s} .
\label{eq:Madelung}
\ee
Then, the real and imaginary parts of the Schr\"odinger equation (\ref{eq:Schrodinger}) give
\ba
&& \dot\rho + \nabla \cdot \left( \rho \frac{\nabla s}{m} \right) = 0 ,
\label{eq:continuity-s} \\
&& \frac{\dot s}{m} + \frac{(\nabla s)^2}{2 m^2} = - ( \Phi_{\rm N} + \Phi_{\rm I} ) ,
\label{eq:Euler-s}
\ea
while the nonrelativistic self-interaction potential reads
\be
\Phi_{\rm I} = \frac{\rho}{\rho_a} = \frac{3 \lambda_4 \rho}{4 m^4} .
\label{eq:Phi_I-rho}
\ee
Defining the curl-free velocity field $\vec v$ by
\be
{\vec v} = \frac{\nabla s}{m} ,
\label{eq:v-def}
\ee
Eqs.(\ref{eq:continuity-s})-(\ref{eq:Euler-s}) give the usual continuity and Euler equations,
\ba
&& \dot\rho + \nabla \cdot ( \rho \vec v ) = 0 ,  \label{eq:continuity} \\
&& \dot {\vec v} + (\vec v \cdot \nabla) \vec v = - \nabla ( \Phi_{\rm N}
+ \Phi_{\rm I} ) . \label{eq:Euler}
\ea
Thus, in the nonrelativistic regime, we can go from the Klein-Gordon equation to the
Schr\"odinger equation and next to a hydrodynamical picture.
In the Hamilton-Jacobi and Euler equations (\ref{eq:Euler-s}) and (\ref{eq:Euler})
we have neglected the quantum pressure term
\be
\Phi_{\rm Q} = - \frac{\nabla^2 \sqrt\rho}{2 m^2 \sqrt\rho} .
\label{eq:Phi_Q-def}
\ee
This is because in this paper we focus on the regime associated with the condition
(\ref{eq:-not-FDM-soliton}) below, where the self-interaction dominates over the quantum pressure.
Then, wavelike effects, such as interference patterns, are negligible.
However, the dynamics remain different from that of CDM particles because of the
self-interaction.

\subsection{Static equilibrium: soliton around a BH}

In contrast with CDM, the pressure due to the self-interaction allows for the formation of static equilibrium configurations with zero velocities
\cite{Chavanis:2011zi,Riotto:2000kh,Harko:2011jy}, which are sometimes called Bose-Einstein condensates or boson stars.
In the more familiar FDM case, such static solutions where gravity is balanced by the quantum pressure (\ref{eq:Phi_Q-def}), rather than by the self-interaction
(\ref{eq:Phi_I-psi}), are often called solitons \cite{Schive:2014hza,Marsh:2015wka,Hui:2016ltb}
and correspond to a bound ground state of the linear Schr\"odinger equation in the Newtonian gravitational potential.
In our case, the self-interaction adds an explicit nonlinearity to the Schr\"odinger equation, through the self-interaction potential $\Phi_{\rm I}$
in Eq.(\ref{eq:Schrodinger}), in addition to the self-gravity included in the
Newtonian potential $\Phi_{\rm N}$.
As we have in mind extended scalar clouds, which may reach galactic size
as for the FDM scenario, rather than compact objects, we call these hydrostatic
equilibrium solitons as in the FDM case, rather than boson stars.
They are again bound ground states of the Schr\"odinger equation
(\ref{eq:Schrodinger}), where the full potential now reads
$\Phi = \Phi_{\rm N} + \Phi_{\rm I}$.
As for FDM, this is actually a nonlinear equation of motion, because of the self-gravity in $\Phi_{\rm N}$ and of the dependence of the self-interaction potential
$\Phi_{\rm I}$ on $\rho=m|\psi|^2$.
From Eq.(\ref{eq:Euler}), the equation of hydrostatic equilibrium reads
\be
\nabla (\Phi_{\rm N} + \Phi_{\rm I}) = 0 ,
\label{eq:hydro-eq}
\ee
which we integrate as
\be
\Phi_{\rm N} + \Phi_{\rm I} = \alpha , \;\;\; \mbox{with} \;\;\; \alpha =
\Phi_{\rm N}(R_{\rm sol}) .
\label{eq:hydro-alpha}
\ee
Here we have introduced the radius $R_{\rm sol}$ of the spherically symmetric soliton, where
the density is zero and hence $\Phi_{\rm I} = 0$, which determines the value of
the integration constant $\alpha$.
The Newtonian gravitational potential is given by the sum of the contributions from
the central BH and from the scalar-cloud self-gravity,
\be
\Phi_{\rm N} = \Phi_{\rm BH} + \Phi_{\rm sg} ,
\ee
with
\be
\Phi_{\rm BH} = - \frac{{\cal G} M_{\rm BH}}{r} = - \frac{r_s}{2 r} , \;\;\;
\nabla^2 \Phi_{\rm sg} = 4\pi {\cal G} \rho ,
\label{eq:Phi_N}
\ee
where $r_s = 2 {\cal G} M_{\rm BH}$ is the Schwarzschild radius of the BH of mass
$M_{\rm BH}$. Taking the divergence of Eq.(\ref{eq:hydro-eq}), using Eqs.(\ref{eq:Phi_N})
and (\ref{eq:Phi_I-rho}) and looking for a spherically symmetric solution, we obtain
\be
\frac{d^2\Phi_{\rm I}}{dr^2} + \frac{2}{r} \frac{d\Phi_{\rm I}}{dr} + \frac{1}{r_a^2}
\Phi_{\rm I} = 0 , \;\;\; \mbox{with} \;\; r_a= \frac{1}{\sqrt{4\pi{\cal G} \rho_a}} ,
\label{eq:ra-rhoa}
\ee
where $r_a$ was also defined in Eq.(\ref{eq:ra-def}).
Introducing the dimensionless radius $x = r/r_a$, we recover the differential equation
satisfied by spherical Bessel functions of order zero. Thus,
$\Phi_{\rm I} = a \, j_0(x) + b \, y_0(x)$. At small radii, the gravitational potential
is dominated by the BH and from Eq.(\ref{eq:hydro-alpha}) we obtain
$\Phi_{\rm I} \simeq r_s/(2r)$. This determines the integration constant $b$,
and we can write the solution for the density $\rho$ in the nonrelativistic regime as
\be
r \gg r_s : \;\;\; \rho(r) = \rho_0 \frac{\sin(r/r_a)}{(r/r_a)} + \rho_a \frac{r_s}{2 r_a}
\frac{\cos(r/r_a)}{(r/r_a)} .
\label{eq:rho-sin-cos}
\ee
The first term dominates at large radii, where the gravitational potential is mostly given
by the soliton self-gravity, while the second term dominates at small radii, where the
gravitational potential is mostly due  to the BH. This transition radius $r_{\rm sg}$ is
typically much smaller than the size of the soliton $R_{\rm sol}$,
and much greater than the Schwarzschild radius,
\be
R_{\rm sol} \simeq \pi r_a , \;\;\; r_{\rm sg} = r_s \frac{\rho_a}{\rho_0} ,
\;\;\; r_s \ll r_{\rm sg} \ll R_{\rm sol} .
\label{eq:r_sg-def}
\ee
Then, far inside the soliton we have
\be
r_s \ll r \ll r_{\rm sg}^{1/3} r_a^{2/3} : \;\;\;  \rho = \rho_0 + \rho_a \frac{r_s}{2 r}
\; .
\label{eq:rho-intermediate}
\ee
In terms of the fields $\psi$ and $\phi$ this static soliton reads
\be
\psi = \sqrt{\frac{\rho}{m}} e^{-i \alpha m t}  , \;\;\; \phi = \frac{\sqrt{2\rho}}{m}
\cos[(1+\alpha) m t ] ,
\label{eq:psi-phi-soliton}
\ee
as the phase $s$ reads $s=-\alpha m t$.

In the case of FDM, where the soliton can reach kpc size, numerical simulations
\citep{Schive:2014hza,Mocz:2017wlg} show that outside this core the scalar field is out
of equilibrium, with large density fluctuations and a mean falloff that follows the NFW profile
\citep{Navarro:1995iw} found for CDM.
We expect a similar behavior for SFDM, in cases where there is a unique soliton
of kpc size inside galaxies.
However, in this paper we also consider scenarios with much smaller values of $r_a$,
where there could be many scalar clouds of smaller size in the galaxy.
In any case, using the hierarchy of scales (\ref{eq:r_sg-def}), we do not specify here the
dark matter profile beyond the soliton radius. As we shall find in Sec.~\ref{sec:nonlinear-flow},
the interaction between the BH and the scalar cloud is governed by radii $r \lesssim r_{\rm sg}$,
that is, radii where the BH gravity is subdominant, and  do not significantly contribute to the
accretion and the dynamical friction of the BH. In contrast with the collisionless case,
there is no infrared divergence and our results do not depend on the dynamics near the
scalar cloud border or beyond.

Therefore, in our derivation of the scalar flow via the relativistic generalization
(\ref{eq:KG-phi-2}) below, the boundary condition at ``large radius''
will actually be the hydrostatic profile within the bulk of the soliton,
at $r_{\rm sg} \ll r \ll R_{\rm sol}$.
What happens at $r \gtrsim R_{\rm sol}$ is beyond the scope of this paper and irrelevant
to the BH dynamics that we investigate here.

\subsection{Moving soliton}

If there is no BH, Galilean invariance ensures that the equilibrium solution
$\rho_{\rm eq}(r)$ also maps to a solution that moves at the uniform velocity
$\vec v_0$, $\rho_{v_0}(\vec r,t) = \rho_{\rm eq}(\vec r - \vec v_0 t)$.
The phase $s$ now reads  $s= - (\alpha + v_0^2/2) m t + m v_0 z$, for a velocity along
the $z$-axis, and the scalar field $\phi$ becomes
\be
\phi = \frac{\sqrt{2\rho}}{m} \, \cos[  (1+\alpha + v_0^2/2) m t - m v_0 z ] .
\label{eq:phi-large-r}
\ee

We now consider the case of a BH moving with the velocity $-\vec v_0$ through the soliton,
or equivalently of a soliton moving at the velocity $\vec v_0$ with respect to a
motionless BH. Neglecting the gravitational backreaction of the scalar cloud, which in
the static case amounts to the dressing of the BH potential by the factor $\cos(r/r_a)$,
and focusing on scales far inside the soliton, we take for the density $\rho$,
the self-interaction potential and the total gravitational potential the expression
(\ref{eq:rho-intermediate}),
\be
r \gg r_s : \;\;\; \rho = \rho_0 + \rho_a \frac{r_s}{2r} , \;\;\; \Phi_{\rm I} =
\frac{\rho}{\rho_a} , \;\;\; \Phi_{\rm N} = \alpha - \Phi_{\rm I} .
\label{eq:boundary-large-radius}
\ee
Together with $\vec v = \vec v_0$, this sets the boundary conditions at large radii.

\subsection{Parameter space}

\begin{figure}
\begin{center}
\includegraphics[width=\columnwidth]{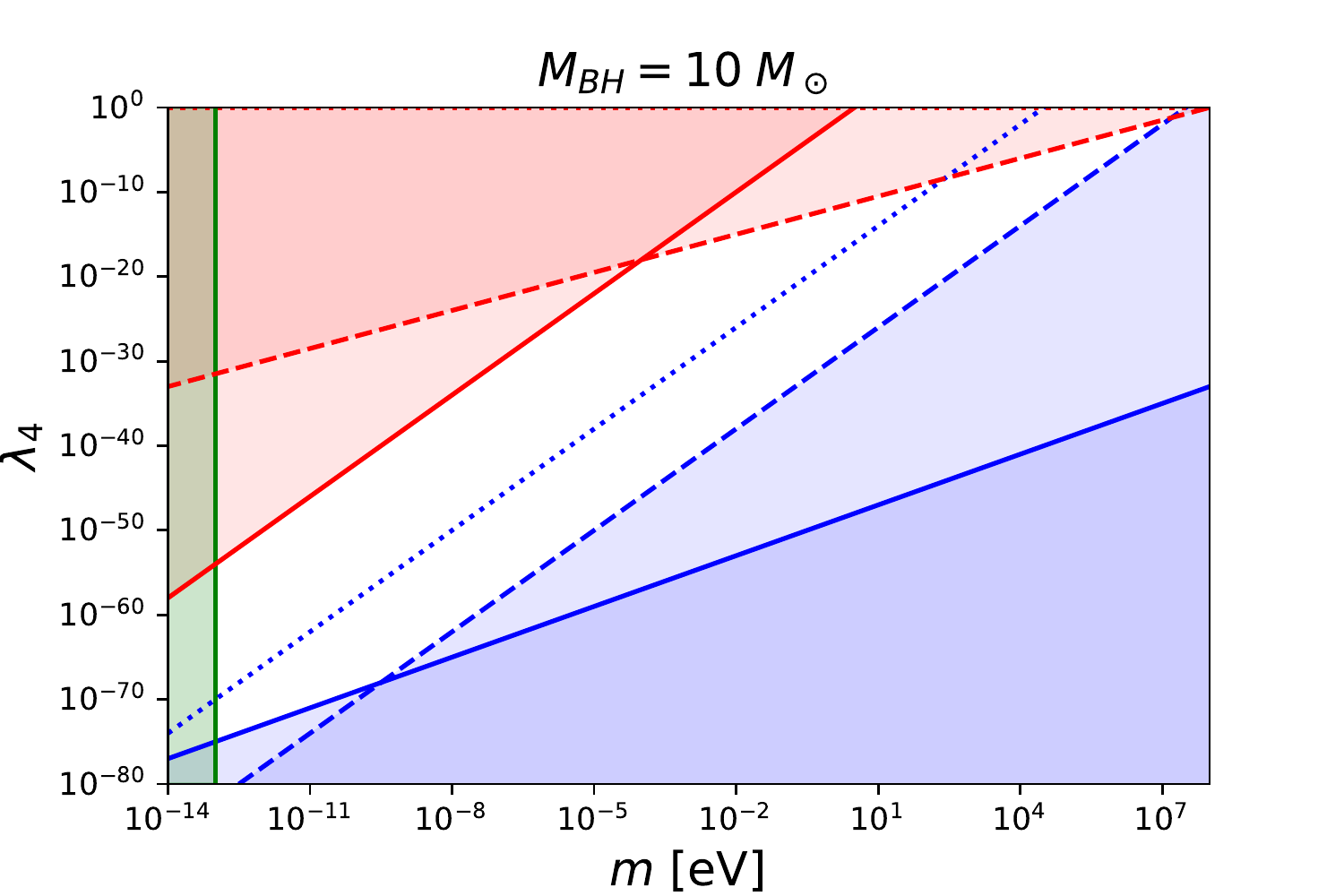}
\includegraphics[width=\columnwidth]{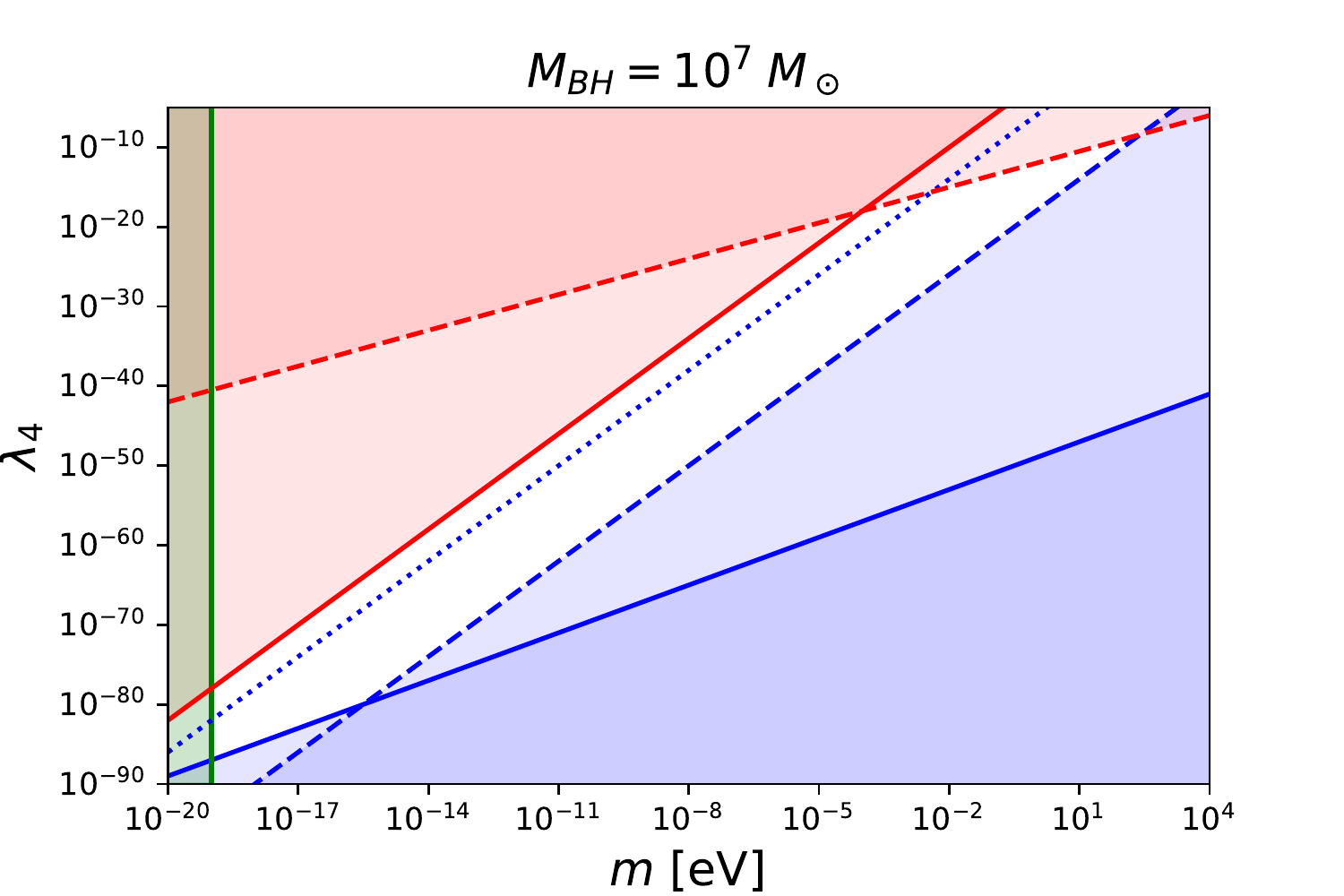}
\end{center}
\caption{
Domain in the parameter space $(m,\lambda_4)$ where our derivations apply, for a BH of mass
$10 \, M_\odot$ (upper panel) and $10^7 M_\odot$ (lower panel). The region in white is the allowed part of the parameter space. 
}
\label{fig_m-lambda4}
\end{figure}

Before we investigate the details of the scalar flow around the BH in the following sections,
we first review the constraints on the scalar parameters $m$ and $\lambda_4$ and
the regime where our computation applies.
We actually have a third parameter, the density $\rho_0$ in the bulk of the soliton.
It gives the transition radius $r_{\rm sg}$ of Eq.(\ref{eq:r_sg-def}),
the values in the bulk of the soliton of the self-interaction potential $\Phi_{\rm I 0}$ from
Eq.(\ref{eq:Phi_I-rho}) and of the scalar-field gravitational potential $\Phi_{\rm sg 0}$.
It also determines the sound speed $c_s^2$, as seen in Eqs.(\ref{eq:k2-infinity}) and
(\ref{eq:vs-def}) below, which is also the typical velocity scale in the soliton,
\be
\frac{\rho_0}{\rho_a} \sim \frac{r_s}{r_{\rm sg}} \sim \Phi_{\rm I 0} \sim c_s ^2 \sim v^2 \lesssim 1 .
\label{eq:v2-rho0-cs2}
\ee

Throughout this paper, we focus on the regime $r_a \gg \lambda_{\rm dB}$,
where the de Broglie wavelength $\lambda_{\rm dB} = 2\pi /(mv)$ sets the scale where
the quantum pressure comes into play, i.e. interference effects associated with the wavelike
behaviors that arise from the Klein-Gordon or Schr\"odinger equations.
At the scale of the scalar cloud, this condition reads
\be
\lambda_4 \gg \frac{m^2}{M_{\rm Pl}^2 v^2} , \;\;\; \mbox{hence} \;\;
\lambda_4 \gg 10^{-55} \, v^{-2} \left(\frac{m}{1\,\rm eV}\right)^2 .
\label{eq:-not-FDM-soliton}
\ee
More generally, from Eqs.(\ref{eq:Phi_I-rho}) and (\ref{eq:Phi_Q-def}), the condition for the
self-interaction $\Phi_{\rm I}$ to dominate over the quantum pressure $\Phi_{\rm Q}$
reads $\lambda_4 \rho/m^4 \gg 1/(m^2 r^2)$, that is, $\lambda_4 \gg m^2/(\rho r^2)$.
Within the equilibrium scalar cloud, where $v^2 \sim \Phi_{\rm N} \sim \rho r^2/M_{\rm Pl}^2$,
we recover Eq.(\ref{eq:-not-FDM-soliton}).
Close to the Schwarzschild radius $r_s$, the scalar field density is of the order of
$\rho_a \sim m^4/\lambda_4$ \citep{Brax:2019npi} and we obtain instead the condition
\be
m \gg \frac{M_{\rm Pl}^2}{M_{\rm BH}} , \;\;\; \mbox{hence} \;\;\;
m \gg 10^{-12} \left( \frac{M_{\rm BH}}{1 \, M_{\odot}} \right)^{-1} {\rm eV} .
\label{eq:not-FDM-BH}
\ee
Therefore, our computation applies to stellar-mass BHs for $m \gg 10^{-12} \, {\rm eV}$
and to supermassive BHs for $m \gg 10^{-18} \, {\rm eV}$.

The regime $\lambda_4=0$, where self-interactions are negligible, is the Fuzzy Dark Matter
scenario \cite{Hui:2016ltb}, where gravity can be balanced by the quantum pressure
at scale $\lambda_{\rm dB}$.
For galactic halos, with $v\simeq 10^{-3}$, to reach $\lambda_{\rm dB} \sim 1$ kpc
so that departures from CDM can be observed on galactic scales
and be relevant for the core-cusp problem gives $m \sim 10^{-22}$ eV.
However, this is in tension with Lyman-$\alpha$ forest constraints
\citep{Irsic:2017yje, Rogers:2020ltq} and the analysis of galactic rotation curves
\citep{Bar:2021kti}, which require $m \gtrsim 10^{-20}$ eV.

In the regime (\ref{eq:-not-FDM-soliton}), where the self-interaction dominates
over the quantum pressure,
departures from CDM on galactic scale, and possible impacts on the core-cusp problem,
can be obtained for a large range of masses, thanks to the additional parameter $\lambda_4$,
which is related to the characteristic scale $r_a$ of Eqs.(\ref{eq:ra-def}) and
(\ref{eq:ra-rhoa}) \citep{Brax:2019fzb}
\be
\lambda_4 \simeq \left(\frac{r_a}{20 \, \rm kpc}\right)^2 \left( \frac{m}{1 \, \rm eV}\right)^4 .
\label{eq:lambda4-ra}
\ee
However, in this paper we do not require the formation of scalar clouds/solitons of
galactic size, relevant for the core-cusp problem.
We consider the more general case of SFDM, independently of its possible impact on
$\Lambda$CDM galactic-scale tensions, where the cloud size $r_a$ may range from astrophysical
to galactic size. Then, assuming such scalar clouds form and include stellar or BH systems,
in a fashion similar to molecular clouds, we investigate the accretion rate and the dynamical friction
of a BH inside such a cloud.

In this paper, to simplify the computation and the boundary condition at large distance, we assume
that $r_a \gg r_{\rm sg}$, that is, the cloud extends beyond the transition radius.
This gives
\be
r_a \gg r_{\rm sg} : \;\;\; \lambda_4 \gg
\left( \frac{M_{\rm BH} m^2}{M_{\rm Pl}^3 v^2} \right)^2 ,
\ee
and hence,
\be
\lambda_4 \gg 10^{-20} \left( \frac{v}{10^{-3}} \right)^{-4}
\left( \frac{M_{\rm BH}}{1 \, M_{\odot}} \right)^2  \left( \frac{m}{1 \, \rm eV}\right)^4 ,
\label{eq:lambda4-Rsol-rsg}
\ee
where $v^2 \sim \Phi_{\rm I} \sim  \Phi_{\rm N}$ is the typical orbital velocity in the gravitational
potential well of the scalar cloud, as in (\ref{eq:v2-rho0-cs2}).
However, this is not a critical constraint and it would be sufficient to require that the cloud extends
much beyond the BH horizon.
In this case, $r_s \ll r_a \ll r_{\rm sg}$, the self-gravity regime is never reached and the profile
remains dominated by the second term in Eq.(\ref{eq:rho-sin-cos}).
Thus, the scalar cloud radius is now $R_{\rm sol} \simeq \pi r_a/2$, again of the order of $r_a$.
This gives the constraint
\be
r_a \gg r_s : \;\;\; \lambda_4 \gg \left( \frac{M_{\rm BH} m^2}{M_{\rm Pl}^3} \right)^2 ,
\ee
and hence,
\be
\lambda_4 \gg 10^{-32} \left( \frac{M_{\rm BH}}{1 \, M_{\odot}} \right)^2
\left( \frac{m}{1 \, \rm eV}\right)^4 .
\label{eq:lambda4-Rsol-rs}
\ee
This would slightly change the boundary condition but the main steps of our derivation, especially
the form (\ref{eq:phi-cn-def}) of the solution, remain valid. However, the dynamical friction
$F_z$ will be decreased because of the smaller size of the scalar cloud, which is smaller than
the radius $r_{\rm sg}$ where the contribution to $F_z$ otherwise peaks, as found
in Sec.~\ref{sec:friction} below.

Observations of the merging of clusters provide the upper bound
$\sigma/m \lesssim 1 \, {\rm cm^2/g}$ for the cross-section $\sigma$, which gives \cite{Brax:2019fzb}
\be
\lambda_4 \lesssim 10^{-12} (m/1\,{\rm eV})^{3/2}.
\label{eq:lambda4-clusters}
\ee
The requirement that $V_{\rm I} \ll V$ since matter-radiation equality (so that the scalar field
behaves like DM) gives $\lambda_4 \lesssim (m/1\,{\rm eV})^4$, which is automatically verified
for $r_a < 20$ kpc from Eq.(\ref{eq:lambda4-ra}).
For the classical description of the scalar field to be valid, the occupation number
$N \simeq (\rho/m) \lambda_{\rm dB}^3 \sim \rho / (m^4 v^3)$ must be much greater than unity.
As we shall see below in Eqs.(\ref{eq:Phi_I-rho}) and (\ref{eq:hydro-alpha}), at equilibrium in the
scalar cloud the balance between gravity and the self-interaction pressure yields
$\Phi_{\rm N} \sim \Phi_{\rm I} \sim \rho\lambda_4/m^4$. Together with $v^2 \sim \Phi_{\rm N}$,
for the typical orbital velocity in the gravitational potential well,
this gives
\be
N \sim 1/(\lambda_4 v) \gg 1 , \;\;\; \mbox{hence} \;\;\; \lambda_4 \ll v^{-1} .
\ee
With $v \lesssim 1$, the classical approximation applies as long as $\lambda_4 \ll 1$.

We will solve the Klein-Gordon equation in curved spacetime in the large scalar mass limit
(\ref{eq:phi-cn-def}), where $m$ is much greater than typical spatial gradients and frequencies.
This requires in particular that $m \gg  1/r_s$, where $r_s$ is the BH horizon,
\be
m \gg \frac{1}{r_s} , \;\;\; \mbox{hence} \;\;\; m \gg \frac{M_{\rm Pl}^2}{M_{\rm BH}} .
\label{eq:m-gg-1-over-rs}
\ee
We recover the same condition as Eq.(\ref{eq:not-FDM-BH}), which ensured that wave effects
are negligible and that we are far from the FDM regime.

The domain in the parameter space $(m,\lambda_4)$ where our derivations apply
corresponds to the central white area in Fig.~\ref{fig_m-lambda4}, for the cases of a BH of mass
$10 \, M_\odot$ (upper panel) and $10^7 M_\odot$ (lower panel).
The scalar mass $m$ is bounded from below by (\ref{eq:not-FDM-BH}), shown by the vertical green
solid line on the left. At low $m$, the coupling $\lambda_4$ is bounded from above by
(\ref{eq:lambda4-ra}), where we take $r_a \leq 2 \, {\rm kpc}$, shown by the upper red solid line.
At high $m$, $\lambda_4$ is bounded from above by (\ref{eq:lambda4-clusters}), shown by the
upper red dashed line.
At low $m$, $\lambda_4$ is bounded from below by (\ref{eq:-not-FDM-soliton}),
where we take $v=10^{-3}$, shown by the lower blue solid line.
At high $m$, $\lambda_4$ is bounded from below by (\ref{eq:lambda4-Rsol-rs}), shown by the
lower blue dashed line.
The condition (\ref{eq:lambda4-Rsol-rsg}) is shown by the blue dotted line for
$v=10^{-3}$. As noticed above, in the following we assume that we are above this threshold
to apply the boundary condition at large distance, but this is not critical as the form
(\ref{eq:phi-cn-def}) of the solution still applies provided we are above the
blue dashed line (\ref{eq:lambda4-Rsol-rs}).
Overall, this mostly gives a diagonal band $\lambda_4 \sim m^4$ in the space $(m,\lambda_4)$,
with for $M_{\rm BH} = 10 \, M_\odot$,
\be
10^{-12} \lesssim m \lesssim 10^7 {\rm eV} , \;\;
10^{-75 } \lesssim \lambda_4 \lesssim 10^{-3}  ,
\ee
and for $M_{\rm BH} = 10^7 \, M_\odot$,
\be
10^{-19} \lesssim m \lesssim 10 \, {\rm eV} , \;\;
10^{-85 } \lesssim \lambda_4 \lesssim 10^{-10}  .
\ee

\section{Relativistic regime}
\label{sec:relativistic_regime}

\subsection{Isotropic metric}

As we neglect the gravitational backreaction of the scalar cloud, we consider the spherically symmetric metric associated with a BH at the center
of a large soliton, matching Eq.(\ref{eq:boundary-large-radius}) at large radii. Moreover we consider
a non-rotating BH and to simplify the matching with the usual Newtonian gauge on large scales,
we work with the isotropic radial coordinate $r$ and time $t$, so that the static spherically symmetric
metric can be written in the isotropic form
\be
ds^2 = - f(r) \, dt^2 + h(r) \, (dr^2 + r^2 \, d\vec\Omega^2) .
\label{eq:ds2-def}
\ee
At large radii in the weak-gravity regime, far beyond the Schwarzschild radius, we have
\be
f = 1 + 2 \Phi_{\rm N} , \;\;\; h = 1 - 2 \Phi_{\rm N} ,
\ee
with
\be
\Phi_{\rm N} = \alpha - \Phi_{\rm I} = \alpha - \frac{\rho_0}{\rho_a} - \frac{r_s}{2 r} ,
\ee
in agreement with (\ref{eq:boundary-large-radius}).
The first two terms in the last expression correspond to the scalar-cloud self-gravity
while the last term is the BH gravitational potential.
At smaller scales where the BH gravity dominates, far inside the transition radius
$r_{\rm sg}$, the isotropic metric functions $f(r)$ and $h(r)$ read
\ba
\frac{r_s}{4} < r \ll r_{\rm sg} : && f(r) = \left( \frac{1-r_s/(4r)}{1+r_s/(4r)}
\right)^2 , \nonumber \\
&& h(r) = (1+r_s/(4r))^4 .
\label{eq:f-h-def}
\ea
In these coordinates, the BH horizon is located at radius $r=r_s/4$.

\subsection{Equations of motion}

\subsubsection{Klein-Gordon equation}

In the metric (\ref{eq:ds2-def}), the Klein-Gordon equation reads
\be
\frac{\partial^2\phi}{\partial t^2} - \sqrt{\frac{f}{h^3}} \nabla \cdot ( \sqrt{f h}
\nabla \phi ) + f \frac{\partial V}{\partial \phi} = 0 .
\label{eq:KG-phi-1}
\ee
As the metric is spherically symmetric and the uniform velocity at large distance is
parallel to the $z$-axis, $\vec v = \vec v_0 = v_0 \, \vec e_z$, the system is axisymmetric
around the $z$-axis. Therefore, the Klein-Gordon equation (\ref{eq:KG-phi-1}) reads
in spherical coordinates as
\ba
&&  \!\!\! \frac{\partial^2\phi}{\partial t^2} - \sqrt{\frac{f}{h^3}} \frac{1}{r^2}
\frac{\partial}{\partial r} \left[ r^2 \sqrt{f h} \frac{\partial\phi}{\partial r} \right] - \frac{f}{h r^2 \sin\theta} \frac{\partial}{\partial\theta} \left[ \sin\theta
\frac{\partial\phi}{\partial\theta} \right] \nonumber \\
&&+ f m^2 \phi + f \lambda_4 \phi^3 = 0 .
\label{eq:KG-phi-2}
\ea
As in \cite{Brax:2019npi}, we note that the cubic nonlinearity is of the same type as for
the Duffing equation \cite{Kovacic-2011}. This allows us to look for a solution of the form
\be
\phi = \phi_0(r,\theta) \, {\rm cn}[ \omega(r,\theta) t - {\bf K}(r,\theta)
\beta(r,\theta), k(r,\theta) ] ,
\label{eq:phi-cn-def}
\ee
where we denoted ${\rm cn}(u,k)$ the Jacobi elliptic function \cite{Gradshteyn1965,Byrd-1971}
of argument $u$, modulus $k$, and period $4 {\bf K}$, where ${\bf K}(k)$ is the complete elliptic
integral of the first kind, defined by
${\bf K}(k) = \int_0^{\pi/2} \, d\theta/\sqrt{1-k^2\sin^2\theta}$ for $0 \leq k <1$
\cite{Gradshteyn1965,Byrd-1971}.
We also defined ${\bf K}(r,\theta) \equiv {\bf K}[k(r,\theta)]$.
Equation (\ref{eq:phi-cn-def}) is the leading-order approximation
in the limit $m \to \infty$, where spatial gradients of the amplitude $\phi_0$ and
the modulus $k$ are much below $m$, while both $\omega$ and $\beta$ are of the order of $m$.
The amplitude $\phi_0$, the angular frequency $\omega$, the phase $\beta$ and the modulus
$k$ are slow functions of space.

Thus, this is a generalization of nonrelativistic expressions such as (\ref{eq:phi-large-r}),
where the usual trigonometric functions are replaced by the Jacobi elliptic function because
of the strong cubic nonlinearity.
We recover the nonrelativistic regime for small modulus $k$, as ${\rm cn}(u,0) = \cos(u)$.
Therefore, the modulus $k$ measures the deviation from harmonic oscillations and from
the nonrelativistic limit.

In contrast with the study of radial accretion presented in \cite{Brax:2019npi}, because of
the incoming velocity $\vec v_0=v_0 \, \vec{e}_z$ at large distance, the scalar-field configuration
is no longer spherically symmetric but only axi-symmetric.
This implies that the amplitude $\phi_0$, the angular frequency $\omega$,
the phase $\beta$ and the modulus $k$ depend on both the radial distance $r$ and
the angle $\theta$ with respect to the $z$-axis.

To ensure that spatial gradients do not increase with time, the oscillations of the scalar field
at different locations must be synchronized.
This means that the function $\omega(r,\theta)$ is set by the modulus
$k(r,\theta)$ according to
\be
\omega(r,\theta) = \frac{2{\bf K}(r,\theta)}{\pi} \, \omega_0 ,
\label{eq:omega-omega0}
\ee
with a constant fundamental frequency $\omega_0$ that will be set by the boundary conditions.

At leading order in the large-$m$ limit, the spatial derivatives read
\ba
&& \frac{\partial^2\phi}{\partial r^2} = \phi_0 \left( {\bf K}
\frac{\partial\beta}{\partial r} \right)^2 {\rm cn}'' + \dots , \\
&& \frac{\partial^2\phi}{\partial \theta^2} = \phi_0 \left( {\bf K}
\frac{\partial\beta}{\partial \theta} \right)^2 {\rm cn}'' + \dots ,
\ea
where the dots stand for subleading terms and we denoted
${\rm cn}'' = \frac{\partial^2{\rm cn}}{\partial u^2}$.
Substituting into the Klein-Gordon equation (\ref{eq:KG-phi-2}) and using the differential equations  satisfied by the Jacobi elliptic functions,
${\rm cn}'' = (2 k^2-1) {\rm cn} - 2 k^2 {\rm cn}^3$,
we obtain the two conditions
\ba
&& (\nabla \beta)^2 = \frac{h}{f} \left( \frac{2 \omega_0}{\pi} \right)^2
- \frac{h m^2}{(1-2k^2) {\bf K}^2} ,
\label{eq:beta-1}
\\
&& \frac{\lambda_4 \phi_0^2}{m^2} = \frac{2 k^2}{1-2k^2}
\label{eq:lambda4-k}
\ea
which will be interpreted in the following.

\subsubsection{Conservation equation}

The system (\ref{eq:beta-1})-(\ref{eq:lambda4-k}) was studied in \cite{Brax:2019npi} for
the case of radial accretion, where $(\nabla \beta)^2$ simplifies as $(d\beta/dr)^2$.
Then, at each radius Eqs.(\ref{eq:beta-1})-(\ref{eq:lambda4-k}) provide $d\beta/dr$ and
$\phi_0$ in terms of the modulus $k$, which remains to be determined.
The profile $k(r)$ then followed from the constraint of a constant flux $F$, to ensure
a steady state solution.
Indeed, the conservation equation $\nabla_\mu T^\mu_0=0$, where $T^\mu_\nu$ is the
energy-momentum tensor of the scalar field, is automatically satisfied at leading order
because Eq.(\ref{eq:phi-cn-def}) is a solution of the Klein-Gordon equation.
At this order, each contribution to $\nabla_\mu T^\mu_0$ is a fast oscillating function
of time with zero mean. However, to go beyond and ensure there are no secular terms, i.e. slowly growing terms in time,  that would violate the
steady state condition, we also require that
$\langle \nabla_\mu T^\mu_0 \rangle =0$, where $\langle \dots \rangle$ is the average
over the oscillations of the solution (\ref{eq:phi-cn-def}). This gives the constraint
\be
\nabla \cdot ( \rho_{\rm eff} \nabla \beta ) = 0 ,
\label{eq:cont-beta}
\ee
with the effective density
\be
\rho_{\rm eff} = \sqrt{fh} \phi_0^2 \omega {\bf K} \langle {\rm cn}'^2 \rangle .
\label{eq:rho_eff}
\ee
Thus, Eqs.(\ref{eq:cont-beta}) and (\ref{eq:beta-1}) generalize to the strong-field
and strong-gravity regimes the continuity and Hamilton-Jacobi equations
(\ref{eq:continuity-s}) and (\ref{eq:Euler-s}), with $\pi \beta/2$ playing the role of the
phase $s$. In the same fashion, they generalize the hydrodynamical continuity and Euler
equations (\ref{eq:continuity}) and (\ref{eq:Euler}), with $\pi \nabla\beta/(2m)$ playing
the role of the curl-free velocity field $\vec v$.
In addition to these continuity and Euler equations, we now have the third equation
(\ref {eq:lambda4-k}).
This is because we now have three fields to determine, the amplitude $\phi_0$ (playing
the role of the density), the phase $\beta$ (playing the role of the velocity potential),
and the modulus $k$.
The latter is coupled to the amplitude by Eq.(\ref {eq:lambda4-k}).
In the nonrelativistic low-amplitude regime this new degree of freedom disappears as we
have $k \to 0$ and the scalar field $\phi$ follows harmonic oscillations, as in
(\ref{eq:phi-large-r}).
In the large-field regime, this new quantity $k(r,\theta)$ determines the amount of
deviation of the nonlinear oscillator from the harmonic oscillator, as described by the
Jacobi elliptic function ${\rm cn}(u,k)$.

\subsection{Low and high velocity branches}
\label{sec:low-high-branches}

In the case of radial accretion, the effective continuity equation (\ref{eq:cont-beta})
can be integrated at once, $F= \rho_{\rm eff} \frac{d\beta}{dr}$, where $F$ is the constant
flux of the scalar field.
Then, using Eqs.(\ref{eq:beta-1})-(\ref{eq:lambda4-k}), we can express
$\rho_{\rm eff} \frac{d\beta}{dr}$ in terms of $k$ and $r$. This gives a condition of the form
$F = {\cal F}(r,k)$.
As seen in \cite{Brax:2019npi}, at each radius $r$, ${\cal F}(r,k)$ seen as a function of
$k$ first increases from 0 at $k=0$, reaches a maximum ${\cal F}_{\rm max}(r)$ at a modulus
$k_{\rm max}(r)$, and then decreases to zero at a modulus $k_+(r)$, turning negative
at higher $k$.
This implies that at each radius $k$ must be in the range $0 \leq k \leq k_+$. Moreover,
if $F > {\cal F}_{\rm max}$ there is no solution, whereas if $F < {\cal F}_{\rm max}$
there are two solutions, $k_1 < k_{\rm max} < k_2$.
The solution $k_1$ corresponds to a high-velocity branch (close to free fall) and the
solution $k_2$ to a low-velocity branch (supported by the pressure built by the
self-interactions).
The boundary condition at the horizon selects the high velocity solution, because the
self-interactions cannot prevent the free fall of dark matter into the BH, whereas the boundary condition
at large radius selects the low-velocity branch, to match with the static equilibrium
soliton.
This selects the critical flux $F_c$, given by the minimum over radii of
${\cal F}_{\rm max}(r)$, reached at a critical radius $r_c$.
This enables the smooth connection of the low-velocity branch $k_2$ at $r>r_c$ to the
high-velocity branch $k_1$ at $r<r_c$, the two branches meeting at the radius $r_c$.
This is similar to the hydrodynamical case \cite{Bondi:1952ni,Michel:1972},
which selects the only value of the flux that provides a transonic solution that connects
the subsonic (i.e. low-velocity) branch at large radii with the supersonic
(i.e. high-velocity) branch at low radii.

At radii where the BH gravity is dominant, this gives the radial profile \cite{Brax:2019npi}
\be
r_s \lesssim r \lesssim r_{\rm sg} : \;\;\; \rho \sim \rho_a \frac{r_s}{r} , \;\;\;
v_r \sim - \frac{r_s}{r} ,
\label{eq:radial-profile-rho-vr}
\ee
whereas at larger radii we have
\be
r \gtrsim r_{\rm sg} : \;\;\; \rho \simeq \rho_0 , \;\;\; v_r \sim - \frac{\rho_a r_s^2}{\rho_0 r^2} .
\label{eq:radial-profile-rho-vr-self}
\ee

In the axi-symmetric case that we consider in this paper, we cannot integrate at once
the conservation equation (\ref{eq:cont-beta}), which is now a two-dimensional partial
differential equation.
However, if the homogeneous velocity $v_0$ at large radii is much smaller than the speed
of light, we can expect the flow to become almost radial much before the critical radius
$r_c$, which is typically of the order of the Schwarzschild radius.
Then, we simply match the flow to the radial case at a radius $r_{\rm m} > r_c$,
which provides the inner boundary condition to our system.
This means that the selection of the critical flux $F_c$ and the transition from the
low-velocity to the high-velocity branch can be identified from the radial case computed as in
\cite{Brax:2019npi}. Then, one only needs to solve the system
(\ref{eq:beta-1})-(\ref{eq:cont-beta}) at large radii along the low-velocity branch to complete the analysis.
This is always valid for the subsonic regime studied in this paper, where the relative velocity
$v_0$ is smaller than the effective speed of sound of the scalar-field soliton at large radii. For larger velocities, discontinuities are expected which are left for future work.

\subsection{Boundary condition at large radii}

At very large radii, $k \ll 1$ and the solution (\ref{eq:phi-cn-def}) takes the form
$\phi = \phi_0 \, \cos(\omega_0t - \pi \beta/2)$, as ${\bf K} \simeq \pi/2$.
Comparing with the nonrelativistic solution (\ref{eq:phi-large-r}), we obtain the boundary
conditions
\be
r \to \infty : \;\;\; \phi_0 = \frac{\sqrt{2 \rho}}{m} , \;\;\; \beta = \frac{2}{\pi}
m v_0 z ,
\label{eq:phi0-beta-infinity}
\ee
and the value of the fundamental frequency $\omega_0$,
\be
\omega_0 = (1+\alpha + v_0^2/2) m .
\label{eq:omega0}
\ee
Then, Eq.(\ref{eq:lambda4-k}) gives the asymptotic behavior of $k$,
\be
r \to \infty : \;\;  k^2 = \frac{\lambda_4 \phi_0^2}{2 m^2} = \frac{4\rho}{3\rho_a}
= \frac{4}{3} \Phi_{\rm I} .
\label{eq:k2-infinity}
\ee
The density $\rho$ is given by Eq.(\ref{eq:boundary-large-radius}).

\subsection{Comparison with Fuzzy Dark Matter derivations}
\label{sec:diff-fuzzy-dm}

The behavior of scalar clouds around BH has already been studied, especially in the case
without self-interactions.
For instance, \cite{Unruh1976} considered a free scalar field in the unperturbed Schwarzschild metric
around a BH. This gives rise to a linear Klein-Gordon equation in curved spacetime, which can be expanded
in spherical harmonics. The radial part obeys a linear second-order differential equation,
with coefficients that depend on the radius.
Focusing on the case $1/m \gg r_s$ (i.e., the Compton wavelength is greater than the
Schwarzschild radius), this problem can be solved by splitting the domain into three
regions (close to the BH, intermediate radii, and large radii). In each region one recovers a standard differential equation that
can be solved in terms of known special functions, and by matching at the inner boundaries
one obtains the global solution.

This problem was recently revisited by \cite{Hui:2019aqm}. Looking for spherically symmetric
solutions in the Schwarzschild metric, the authors could express the general solution
in terms of confluent Heun functions. They could then derive several approximate solutions
depending on the hierarchy between the Compton wavelength $1/m$, the Schwarzschild
radius $r_s$, and the self-gravity radius $r_{\rm sg}$.
Our large-mass regime (\ref{eq:m-gg-1-over-rs}), $m \gg 1/r_s$, corresponds to their regime
IV (particle limit), where there is no potential barrier (i.e. no reflection of incoming waves)
and the fluid falls into the BH. At intermediate radii, they obtain in this regime IV,
\be
\mbox{FDM} , \;\;\; r_s \lesssim r \lesssim r_{\rm sg}: \;\;\; \phi \sim r^{-3/4} e^{-imt-i2m\sqrt{r r_s}} .
\label{eq:FDM-phi-exp}
\ee
The power-law exponent $-3/4$ gives rise to a density profile
\be
\mbox{FDM} , \;\;\; r_s \lesssim r \lesssim r_{\rm sg}: \;\;\; \rho \sim r^{-3/2} .
\label{eq:FDM-rho-radial}
\ee
This can be understood as follows \cite{Hui:2019aqm}.
The free-fall velocity onto the BH behaves as $v_r \sim r^{-1/2}$. For a steady state solution,
the flux of matter through a shell of radius $r$, $4\pi r^2 \rho v_r$, must be independent of $r$,
which implies $\rho \sim r^{-2} v_r^{-1} \sim r^{-3/2}$, in agreement with (\ref{eq:FDM-rho-radial}).

These derivations do not apply to our case because of the self-interaction, which adds a cubic
nonlinearity to the Klein-Gordon equation (\ref{eq:KG-phi-2}).
(As compared with \cite{Unruh1976}, we actually consider the opposite limit $1/m \ll r_s$ where the scalar field probes sub-horizon distances,
see Eq.(\ref{eq:m-gg-1-over-rs}), and we take into account the self-gravity of the scalar field
at large radii, where it dominates over the BH gravity and converges to the static soliton
solution.)
To handle the cubic nonlinearity, from the BH horizon to the soliton radius, we precisely
take advantage of the large scalar mass limit to write the solution in the form (\ref{eq:phi-cn-def}).
This separation of scale allows us to consider the scalar field as a cubic oscillator locally,
at each radius, which is exactly solved by the Jacobi elliptic function. The dependence on radius
is then taken into account by the radial dependence of the amplitude $\phi_0$, the
angular frequency $\omega$, the phase $\beta$ and the modulus $k$, and by the conservation
equation (\ref{eq:cont-beta}).
In the radial case, this set of coupled one-dimensional equations can be integrated
as explained in Sec.~\ref{sec:low-high-branches} and in \cite{Brax:2019npi}.
In the nonradial case, as explained in Secs.~\ref {sec:eq_of_motion_phase} and
\ref{sec:flow_SFDM_BH} below, we will use the mapping to familiar hydrodynamical equations
to follow the behavior of the flow at large nonrelativistic radii, where the transition from
the uniform incoming flow at velocity $\vec v_0$ to the radial infall takes place.

Far from the BH horizon, the modulus $k$ is small and the Jacobi elliptic function
(\ref{eq:phi-cn-def}) becomes a cosine at leading order, giving
$\phi \sim \phi_0(r) e^{i\omega(r) t- {\bf K}\beta(r)}$.
Although we recover harmonic oscillations with time, as for the FDM case
(\ref{eq:FDM-phi-exp}), the nonlinearity associated with the self-interaction remains essential.
Indeed, at large radii in the bulk of the soliton the scalar self-gravity remains balanced by the
self-interaction pressure.
On intermediate radii, this additional pressure support slows down the infall and makes
the radial velocity $v_r$ follow the low-velocity branch discussed in
Sec.~\ref{sec:low-high-branches}, instead of the high-velocity branch associated with the free-fall
velocity $v_r \sim r^{-1/2}$ as for FDM. This leads to the different density slope,
$\rho\propto r^{-1}$ in Eq.(\ref{eq:radial-profile-rho-vr}), as compared with $\rho\propto r^{-3/2}$
in the FDM case (\ref{eq:FDM-rho-radial}).

The FDM case has also been considered in a linear theory treatment
\citep{Lancaster:2019mde,Annulli2020},
using the Keplerian gravitational potential $\Phi_{\rm N}=-r_s/(2r)$ for the background and
looking for linear perturbations to the gravitational potential and the scalar field.
However, this is not possible in our case. Indeed, as explained above and in
Sec.~\ref{sec:low-high-branches}, because of the pressure induced by the self-interaction,
the infall is slowed down by a bottleneck near the Schwarzschild radius, which selects the
transonic solution so that the density near the horizon is of the order of $\rho_a$ and
the radial velocity close to the speed of light, $v_r \sim -1$, see Eq.(\ref{eq:radial-profile-rho-vr}).
This selects the accretion rate on the BH and the infalling flux at all radii, by conservation of matter,
as also discussed in Sec.~\ref{sec:accretion} below.
Therefore, the amplitude in the large-radius Newtonian regime is actually set by the boundary
condition at the BH horizon. This requires a fully nonlinear and relativistic treatment
and the global solution cannot be obtained by a perturbative weak-gravity approximation
alone.

\section{Description of the nonlinear velocity flow}
\label{sec:eq_of_motion_phase}

\subsection{Low-\texorpdfstring{$k$}{Lg} regime}

At radii above $r_c$, the modulus $k$ is small, as we already have $k \simeq 0.4$ at $r_c$,
but nonzero.
The gravitational potential $\Phi_{\rm N}$ is also small at radii much beyond the
Schwarzschild radius.
In this regime, we can simplify the system (\ref{eq:beta-1})-(\ref{eq:cont-beta}).
Eqs.(\ref{eq:lambda4-k}) and (\ref{eq:rho_eff}) give
\be
\phi_0^2 = \frac{2 m^2 k^2}{\lambda_4} , \;\;\; \rho_{\rm eff} =
\frac{\pi m^2 k^2}{2 \lambda_4} \omega_0 \propto k^2 ,
\ee
while Eq.(\ref{eq:beta-1}) reads
\ba
\frac{\pi^2(\nabla\beta)^2}{4 m^2} & = & 2 \alpha + v_0^2 - 2 \Phi_{\rm N}
- \frac{3}{2} k^2 \nonumber \\
& = & 2 \frac{\rho_0}{\rho_a} + \frac{r_s}{r} + v_0^2 - \frac{3}{2} k^2 ,
\ea
where we used Eq.(\ref{eq:boundary-large-radius}).
We can check that this is consistent with the boundary conditions
(\ref{eq:phi0-beta-infinity}) and (\ref{eq:k2-infinity}).
Defining the dimensionless radius $\hat r$ and the rescaled phase $\hat\beta$ by
\be
\hat r = \frac{r}{r_s} , \;\;\; \hat\beta = \frac{\pi}{2m r_s} \beta ,
\label{eq:x-hat-beta-def}
\ee
this becomes
\be
(\hat\nabla \hat\beta)^2 = \frac{3}{2} k_0^2 + v_0^2 + \frac{1}{\hat r} - \frac{3}{2} k^2
= \frac{3}{2} \left[ k_+(\hat r)^2 - k^2 \right] ,
\label{eq:beta-hat-k_+}
\ee
where we introduced the limiting value
\be
k_+(\hat r)^2 = k_0^2 + \frac{2}{3} v_0^2 + \frac{2}{3 \hat r} .
\label{eq:k_+_def}
\ee
This provides the upper bound $k_+(\hat r)$ at radius $r$ for the modulus $k$, as the left-hand
side in Eq.(\ref{eq:beta-hat-k_+}) is positive.
As in the radial accretion case presented in \cite{Brax:2019npi}, the low-velocity branch
corresponds to
\be
\mbox{low-velocity branch:} \;\; k \simeq k_+ , \;\;\; v^2 \ll k_+^2 ,
\label{eq:low-radial-velocity}
\ee
where we defined the velocity $\vec v = \hat\nabla \hat\beta$.
Thus, the modulus $k$ is close to the upper bound $k_+$ and the velocity $v$ is much smaller
than the free-fall value of order $1/\hat r$ (in the weak-gravity regime dominated by the
BH gravity).
This is due to the self-interactions, which act as a pressure force that slows down the
collapse towards the BH.
Then, the conservation equation (\ref{eq:cont-beta}) gives
\be
\hat\nabla \cdot (k^2 \hat\nabla \hat\beta) = 0 .
\label{eq:continuity-low-k}
\ee
Using Eq.(\ref{eq:beta-hat-k_+}) this becomes
\be
\hat\nabla \cdot \left[ \left( k_+(\hat r)^2 - \frac{2}{3} (\hat\nabla \hat\beta)^2 \right)
\hat\nabla \hat\beta \right]  = 0 .
\label{eq:flow-k_+-hatbeta}
\ee

At large radii, the constant modulus $k_+^2 \simeq k_0^2 + 2 v_0^2/3$ and the uniform
velocity $\hat\nabla\hat\beta = \vec v_0$ are indeed solutions of this nonlinear equation.
At the radius $r_{\rm m}$ this can be matched to the radial solution, as explained in
Sec.~\ref{sec:low-high-branches}.
Equation (\ref{eq:flow-k_+-hatbeta}) provides a closed partial differential equation
for the phase $\hat\beta$ of the scalar field, which is also the velocity potential.
This gives in turns the scalar field density $\rho$ and the modulus $k$, describing the
departure from harmonic oscillations.

\subsection{Isentropic potential flow}
\label{sec:isentropic}

Equations (\ref{eq:continuity-low-k}) and (\ref{eq:beta-hat-k_+}) coincide with the
steady-state continuity and Bernouilli equations of an isentropic potential flow,
\be
\hat\nabla \cdot (\hat\rho \vec v) = 0 , \;\;\; \frac{v^2}{2} + V + H = 0 ,
\label{eq:continuity-Bernouilli}
\ee
where $\vec v= \hat\nabla \hat\beta$ is the curl-free velocity, with velocity potential
$\hat\beta$, $V(\hat r)$ is the external-force potential, and $H(\hat\rho)$ is the enthalpy,
with the mapping
\be
\hat\rho = \frac{3}{2} k^2 , \;\;\; V(\hat r) = - \frac{3}{4} k_+^2(\hat r) , \;\;\;
H(\hat\rho) = \frac{\hat\rho}{2}  .
\label{eq:rho-V-H}
\ee
This gives for the effective pressure, defined by $dH=d\hat P/\hat\rho$,
\be
\hat P(\hat\rho) = \hat\rho^2/4 ,
\label{eq:P-rho-H}
\ee
with a polytropic exponent $\gamma_{\rm ad}=2$.
From the Bernouilli equation (\ref{eq:continuity-Bernouilli}) the density can be
expressed in terms of the velocity by
\be
\hat\rho = \gamma + \frac{1}{\hat r} - v^2 ,
\label{eq:rho-gamma-v2}
\ee
where we introduced the parameter $\gamma$,
\be
\gamma = \frac{3}{2} k_0^2 + v_0^2 , \;\;\; \mbox{hence} \;\;\;
\frac{3}{2} k_+^2 = \gamma + \frac{1}{\hat r} .
\label{eq:gamma-def}
\ee

\section{Scalar-field flow around the BH}
\label{sec:flow_SFDM_BH}

\subsection{Linear flow (low velocities)}
\label{sec:linear-flow}

At small radii but far above the Schwarzschild radius, dark matter is in the  low-velocity
radial accretion regime as in (\ref{eq:low-radial-velocity}), so that the
term $(\hat\nabla\hat\beta)^2$ is small as compared with $k_+^2$ in
Eq.(\ref{eq:flow-k_+-hatbeta}). At large radii where $\vec v \simeq \vec v_0$,
this is only true if $v_0 \lesssim k_0$, that is, the BH moves with a speed that is smaller
than the speed of sound of the soliton cloud.
We focus on this regime in this paper, and we will study the high-velocity supersonic case
in a companion paper.
Then, it is useful to consider the ``linear flow'' associated with the linearized version of
Eq.(\ref{eq:flow-k_+-hatbeta}),
\be
\hat\nabla \cdot \left[  k_+(\hat r)^2  \hat\nabla \hat\beta \right]  = 0 ,
\label{eq:flow-k_+-hatbeta-linear}
\ee
as this is a good approximation at all radii.
Thanks to the simple form (\ref{eq:k_+_def}) of the kernel $k_+^2$, this linear equation can
be explicitly solved.
The spherical symmetry of $k_+^2$ implies that the angular part of the linear modes can be
expanded in spherical harmonics, which are eigenfunctions of the angular Laplacian.
As we look for axi-symmetric solutions, we only need the modes
$Y_{\ell}^0(\theta,\varphi)$, that is, the Legendre polynomials $P_\ell(\cos\theta)$.
Thus, the independent axi-symmetric modes $G_\ell(x,\theta)$ are
\be
G_\ell(\hat r,\theta) = G_{\ell}(\hat r) \, P_\ell(\cos\theta) ,
\label{eq:Legendre-G_ell}
\ee
with
\be
\frac{d}{d\hat r} \left( \hat r^2 k_+^2 \frac{d G_\ell}{d\hat r} \right) - \ell (\ell+1)
k_+^2 G_\ell = 0 .
\ee
Introducing the characteristic radius $\hat r_\gamma$,
\be
\hat r_\gamma = 1/\gamma ,
\label{eq:r_gamma-def}
\ee
where $\gamma$ was defined in Eq.(\ref{eq:gamma-def}),
we obtain for $\ell=0$ the growing and decaying modes
\be
G_0^+(\hat r) = 1 , \;\;\; G_0^-(\hat r) = \ln\left( 1+\frac{1}{\gamma \hat r}\right) ,
\label{eq:G0_+_-}
\ee
and for $\ell \neq 0$
\ba
&& G_\ell^+(\hat r) = (\gamma \hat r)^{a-\nu} \; _2F_1(a,1-b;1-b+a;-\gamma \hat r) , \;\;\;
\label{eq:Gell_+} \\
&& G_\ell^-(\hat r) = (\gamma \hat r)^{-\nu} \; _2F_1(a,b;c;-1/(\gamma \hat r)) ,
\label{eq:Gell_-}
\ea
with
\ba
&& \nu = \frac{1+\sqrt{1+4\ell(\ell+1)}}{2} , \;\; a = \nu + \sqrt{\nu(\nu-1)} ,
\nonumber \\
&& b=  \nu - \sqrt{\nu(\nu-1)} , \;\; c = 2 \nu .
\ea
These functions have the low-radius behaviors
\ba
&& \hat r \ll \hat r_\gamma : \;\;\; G_0^+(\hat r) = 1, \;\;\; G_0^-(\hat r) \sim
\ln(1/\hat r) , \nonumber \\
&& \ell \neq 0 : \;\;\;  G_\ell^+(\hat r) \sim \hat r^{\sqrt{\ell (\ell+1)}} , \;\;\;
G_\ell^-(\hat r) \sim \hat r^{-\sqrt{\ell (\ell+1)}} , \hspace{0.6cm}
\label{eq:G-small-r}
\ea
and the large-radius behaviors
\ba
&& \hat r \gg \hat r_\gamma : \;\;\; G_0^+(\hat r) = 1, \;\;\; G_0^-(\hat r) \sim \hat r^{-1} ,
\nonumber \\
&& \ell \neq 0 : \;\;\;  G_\ell^+(\hat r) \sim \hat r^{\ell} ,
\;\;\; G_\ell^-(\hat r) \sim  \hat r^{-\ell-1} .
\label{eq:G-large-r}
\ea
As expected, at large radii where $k_+^2$ goes to a constant, we recover at leading order
the modes of the Laplacian.

The boundary condition at large radius is $\vec v \to v_0 \, \vec e_z$, that is,
\be
\hat r \to \infty : \;\;\; \hat\beta = v_0  \hat r  \cos\theta ,
\ee
while the inner boundary condition at $\hat r_{\rm m}$ sets the radial component,
$v_r \simeq v_r(\hat r_{\rm m})$, that is,
\be
\hat r = \hat r_{\rm m} : \;\;\; \frac{\partial\hat\beta}{\partial \hat r} \simeq
v_r^{\rm m} , \;\;\; \frac{\partial\hat\beta}{\partial\theta} \simeq 0 .
\ee
Thus, at the linear level, which we denote by the superscript $L$, the boundary conditions
only generate the monopole and the dipole,
\be
\hat\beta^L = \hat\beta_0^L(\hat r) + \hat\beta_1^L(\hat r) \, \cos(\theta) ,
\label{eq:beta_L-def}
\ee
with
\be
\hat\beta_0^L(\hat r) = \frac{ v_r^{\rm m} } { G_0^{- '} (\hat r_{\rm m}) } G_0^{-}(\hat r) ,
\label{eq:beta0-L}
\ee
and
\ba
&& \hat\beta_1^L(\hat r) = \frac{v_0}{\gamma} \; (\gamma \hat r)^{\sqrt{2}} \;
\frac{\Gamma(-1+\sqrt{2}) \, \Gamma(2+\sqrt{2})}{\sqrt{2} \, \Gamma(1+2\sqrt{2})}
\nonumber \\
&& \times \; _2F_1(2+\sqrt{2},-1+\sqrt{2};1+2\sqrt{2};-\gamma \hat r) .
\label{eq:beta1-2F1}
\ea
From Eq.(\ref{eq:beta0-L}) we find that in the range where the flow is
approximately radial the velocity decreases as $v_r \sim 1/r$. This agrees with the results
obtained in \cite{Brax:2019npi} for the purely radial accretion.
The modulus $k$ also decreases as $k^2 \simeq k_+^2 \simeq 2/(3 \hat r)$, and the density as
$\rho \propto k^2 \propto 1/r$. This gives indeed a constant radial flux,
$F \propto r^2 \rho v_r$, as required for a steady state.

As a numerical example, we can take $\hat r_{\rm m} \sim 10$ and $v_r(\hat r_{\rm  m}) \sim 0.1$,
somewhat beyond the critical radius $\hat r_c$ associated with the transition between
the low- and high-velocity branches. Thus, we obtain at small radii
\ba
\hat r \ll \hat r_\gamma & : & \;\;\; \hat\beta_0^L \sim \ln(1/\hat r) , \;\;\;
\hat\beta_1^L \sim \frac{v_0}{\gamma} (\gamma \hat r)^{\sqrt{2}} , \nonumber \\
&& \hat\beta_0^{L '} \sim - 1/ \hat r  , \;\;\; \hat\beta_1^{L '} \sim v_0
(\gamma \hat r)^{\sqrt{2}-1} ,
\label{eq:beta0-beta1-small-r}
\ea
and at large radii
\ba
\hat r \gg \hat r_\gamma & : & \;\;\; \hat\beta_0^L \sim (\gamma \hat r)^{-1} , \;\;\;
\hat\beta_1^L \sim v_0 \hat r , \nonumber \\
&& \hat\beta_0^{L '} \sim - 1/(\gamma \hat r^2)  , \;\;\; \hat\beta_1^{L '} \sim v_0 .
\ea
Thus, we find that the linear flow becomes radial at a transition radius $\hat r_{\rm t}$
greater than $\hat r_\gamma$ if $v_0 \ll \gamma$, which gives
\be
v_0 \ll k_0^2 : \;\;\; \hat r_{\rm t}= \frac{1}{\sqrt{\gamma v_0}} \gg \hat r_\gamma ,
\label{eq:rt-low-v0}
\ee
where we used $v_0 \ll 1$.
For larger velocities the transition occurs below $\hat r_\gamma$,
\be
k_0^2 \ll v_0 \ll 1 : \;\;\; \hat r_{\rm t}= \gamma^{-1} ( v_0/\gamma )^{-1/\sqrt{2}} \ll
\hat r_\gamma .
\label{eq:rt-high-v0}
\ee
In practice, for relaxed systems we expect $v_0^2 \sim k_0^2$, that is, a squared velocity of the
order of the gravitational potential of the scalar soliton, with $k_0^2 \ll 1$.
This gives $v_0 \sim k_0 \gg k_0^2$. Therefore, the linear flow typically becomes
radial far inside the radius $\hat r_\gamma$.
There, the amplitude of the dipole $\hat\beta_1^L$ has already somewhat decreased
below the large-distance uniform flow $\vec v_0$, as seen from the exponents
in (\ref{eq:beta0-beta1-small-r}).
Thus, the pressure associated with the self-interactions slows down the linear flow
before it becomes radial and accelerates towards the BH.

\subsection{Large-radius expansions}
\label{sec:Mach-large-radii}

It is possible to go beyond the linear-flow approximation (\ref{eq:flow-k_+-hatbeta-linear})
by looking instead for a large-radius expansion.
As shown in the following section, this allows us to explicitly see how the flow goes
from a subsonic to a supersonic regime as the relative velocity $\vec v_0$ becomes
greater than the sound speed.
We also obtain the analytical expressions of the subleading odd corrections,
of order $\hat r^0$, and even corrections, of order $1/\hat r$, to the uniform-flow
potential $\hat\beta_0 = v_0\hat r \cos\theta$.
These explicit large-distance results will also be useful in Secs.~\ref{sec:accretion}
and \ref{sec:friction} to obtain the accretion rate on the BH and its dynamical friction,
as we will see that they can be read from these large-distance expansions.

\subsubsection{Condition for the subsonic regime at large radii}

It is well known that the hydrodynamical Euler equation can lead to discontinuous
solutions with shocks or contact discontinuities. As suggested by the hydrodynamical
mapping (\ref{eq:continuity-Bernouilli}), this will also happen in our case.
As usual, a low velocity $v_0$ will lead to a subsonic and continuous flow at large radii,
whereas a large velocity $v_0$ will lead to a supersonic flow with a bow shock.
(At small radii, close to the Schwarzschild radius, there is always a supersonic high-velocity
region as explained in Sec.~\ref{sec:low-high-branches}).

At large radii, the velocity is close to $\vec v_0$ and
$\hat \beta \simeq v_0 \hat r \cos\theta$. Therefore, we write in this section
\be
\hat\beta = v_0 \hat r  \cos\theta + \delta\hat\beta
\ee
and we linearize the equation of motion (\ref{eq:flow-k_+-hatbeta}) in $\delta\hat\beta$.
This gives
\be
\frac{\partial^2\delta\hat\beta}{\partial\hat x^2} + \frac{\partial^2\delta\hat\beta}
{\partial\hat y^2} + \left( 1 - \frac{4 v_0^2}{3 k_0^2} \right)
\frac{\partial^2\delta\hat\beta}{\partial\hat z^2}
= \frac{2 v_0 \cos\theta}{3 k_0^2 \hat r^2} ,
\label{eq:delta-beta-z}
\ee
where we work in the cartesian coordinates, $\{\hat x, \hat y, \hat z\}$, with
$\vec v_0 = v_0 \, \vec e_z$. The source term is due to the BH gravity, through the
$1/\hat r$ factor in $k_+^2$, which makes the flow depart from the homogeneous flow
$\vec v_0$. Introducing the sound speed $c_s$,
\be
c_s = \frac{\sqrt{3}}{2} k_0 , \;\;\; \mbox{and} \;\;\; c_s^2 = \frac{d\hat P_0}{d\hat\rho_0} ,
\label{eq:vs-def}
\ee
where in the second expression we used Eq.(\ref{eq:P-rho-H}),
the partial differential equation (\ref{eq:delta-beta-z}) changes character,
from elliptic to hyperbolic, at $v_0=c_s$:
\ba
&& v_0 < c_s : \;\;\; \mbox{elliptic subsonic flow} ,  \label{eq:subsonic} \\
&& v_0 > c_s : \;\;\; \mbox{hyperbolic supersonic flow} .  \label{eq:supersonic}
\ea
Note that this linear analysis in the perturbation $\delta\hat\beta$ is different
from the ``linear flow'' studied in Sec.~\ref{sec:linear-flow}.
There, in Eq.(\ref{eq:flow-k_+-hatbeta-linear}), we linearized the equation of motion
(\ref{eq:flow-k_+-hatbeta}) in $\hat\beta$ itself. Thus, we assumed small velocities
everywhere in space as compared with $k_+^2(\hat r)$, which implies $v_0 \ll c_s$ at large
radii.
In contrast, in Eq.(\ref{eq:delta-beta-z}) we study linear perturbations with respect to the
uniform flow $\vec v_0$, which is dominant at large distance, and we no longer assume
that $v_0$ is small.
Therefore, although the analysis (\ref{eq:delta-beta-z}) is now restricted to large radii,
it allows us to study all regimes for $v_0$. In particular, we can see that already at
large distance the cubic nonlinearity in (\ref{eq:flow-k_+-hatbeta}) introduces a richer
behavior than the ``linear flow'' (\ref{eq:beta_L-def}), as the structure of the dynamics
can change from elliptic (as in that low-velocity case) to hyperbolic.

\subsubsection{Smooth flow at large radii for low relative velocity}

In this paper, we focus on the subsonic case (\ref{eq:subsonic}) and we leave the
supersonic case (\ref{eq:supersonic}) to a companion paper.
For such a low relative velocity, $v_0 < c_s$, the partial differential equation
(\ref{eq:delta-beta-z}) is elliptic, as in a subsonic regime, and the flow is smooth.
Introducing the quantity $\mu>0$ with
\be
0 \leq v_0 < c_s : \;\;\; \mu^2 = 1 - v_0^2/c_s^2 ,
\;\;\; 0 < \mu \leq 1 ,
\label{eq:mu-def-low-v0}
\ee
and rescaling coordinates from $\{\hat x, \hat y, \hat z\}$ to $\{\tilde x, \tilde y, \tilde z\}$,
with $\tilde x= \hat x$, $\tilde y= \hat y$, $\tilde z = \hat z / \mu$,
we recover the usual Laplacian in the left-hand side in Eq.(\ref{eq:delta-beta-z}).
Using the Green function of the 3D Laplacian, we obtain the inhomogeneous solution
\be
\delta\hat\beta_{\rm odd} (\hat{\vec x}) = - \frac{v_0}{6\pi k_0^2} \int \!\!
\frac{ d \hat{\vec x}^{\, '} \; \hat z'}
{\hat r'^{\,3} \; \sqrt{ \mu^2 | \hat{\vec x}^{\,'} \!-\! \hat{\vec x} |^2
+ (1 \!-\! \mu^2) (\hat z^{\,'} \!-\! \hat z)^2} }
\label{eq:delta-beta-int-elliptic}
\ee
where we moved back to the coordinates $\{\hat x, \hat y, \hat z\}$.
Performing the integration (e.g., by introducing Feynman parameters as for the computation
of usual Feynman diagrams in particle physics), we obtain
\be
\delta\hat\beta_{\rm odd} (\hat{\vec x}) = \frac{1}{2 v_0}
\ln \left[ \frac { \mu (1+\cos\theta) }{ \cos\theta+\sqrt{\mu^2+(1-\mu^2)\cos^2\theta} }
\right] ,
\label{eq:deltabeta-odd-large-r-low-v0}
\ee
which is odd in $\cos\theta$ and does not depend on $\hat r$.
This gives a vanishing radial velocity and the angular velocity
\be
\delta v_{\theta} (\hat{\vec x}) = \frac{1}{2 v_0 \hat r \sin\theta} \left[
\frac{1}{\sqrt{\mu^2+(1-\mu^2)\cos^2\theta}} - 1 \right] ,
\label{eq:delta-v-theta-odd}
\ee
which is even in $\cos\theta$ and decays as $1/\hat r$.
This result can also be obtained in a much simpler fashion by looking for a solution
of Eq.(\ref{eq:delta-beta-z}) of the form $\delta\hat\beta(\cos\theta)$ that only depends
on the variable $u=\cos\theta$.

The homogeneous solutions of Eq.(\ref{eq:delta-beta-z}) are the solutions
of the Laplace equation in the coordinates $\{\tilde x, \tilde y, \tilde z\}$.
Expanding in spherical harmonics, we obtain the usual growing and decaying modes,
which behave as in (\ref{eq:G-large-r}). The leading-order decaying solution is
thus the monopole $\delta\hat\beta \propto \tilde r^{-1}$.
Going back to the coordinates $\{\hat x, \hat y, \hat z\}$ this gives
\be
\delta\hat\beta_{\rm even} = \frac{B}{\hat r} \left[ \mu^2
+ (1-\mu^2) \cos^2\theta \right]^{-1/2} ,
\label{eq:deltabeta-even-large-r}
\ee
with a normalization factor $B$.
It decays as $1/\hat r$ and is even in $\cos\theta$.
Thus, the even components of the velocity potential decay faster than the odd components
and correspond to a subsubleading correction. It is not necessary to consider the quadratic terms
over $\delta\hat\beta^2$ to obtain this even component because of the partial decoupling
of different parities in the nonlinear equation (\ref{eq:flow-k_+-hatbeta}): the odd term
(\ref{eq:deltabeta-odd-large-r-low-v0}) only generates an odd term at order $1/\hat r^3$.
Therefore, at order $1/\hat r^3$ the even component is fully determined by the linear operator
in the left-hand side of Eq.(\ref{eq:delta-beta-z}), which gives (\ref{eq:deltabeta-even-large-r}).

Expanding $\delta\hat\beta$ in Legendre polynomials, as in
App.~\ref{sec:mode-coupling-large-radii}, we obtain the large-radius behaviors
\be
\hat r \gg \hat r_\gamma : \;\;\; \delta\hat\beta_{2\ell+1} \sim \hat r^0 ,
\;\;\; \delta\hat\beta_{2\ell} \sim \hat r^{-1} ,
\ee
while both odd and even multipoles decay for $v_0 \to 0$ as
\be
v_0 \to 0 : \;\;\; \delta\hat\beta_n \sim v_0^n .
\label{eq:deltabeta_l_v0}
\ee

\begin{figure}
\begin{center}
\includegraphics[width=\columnwidth]{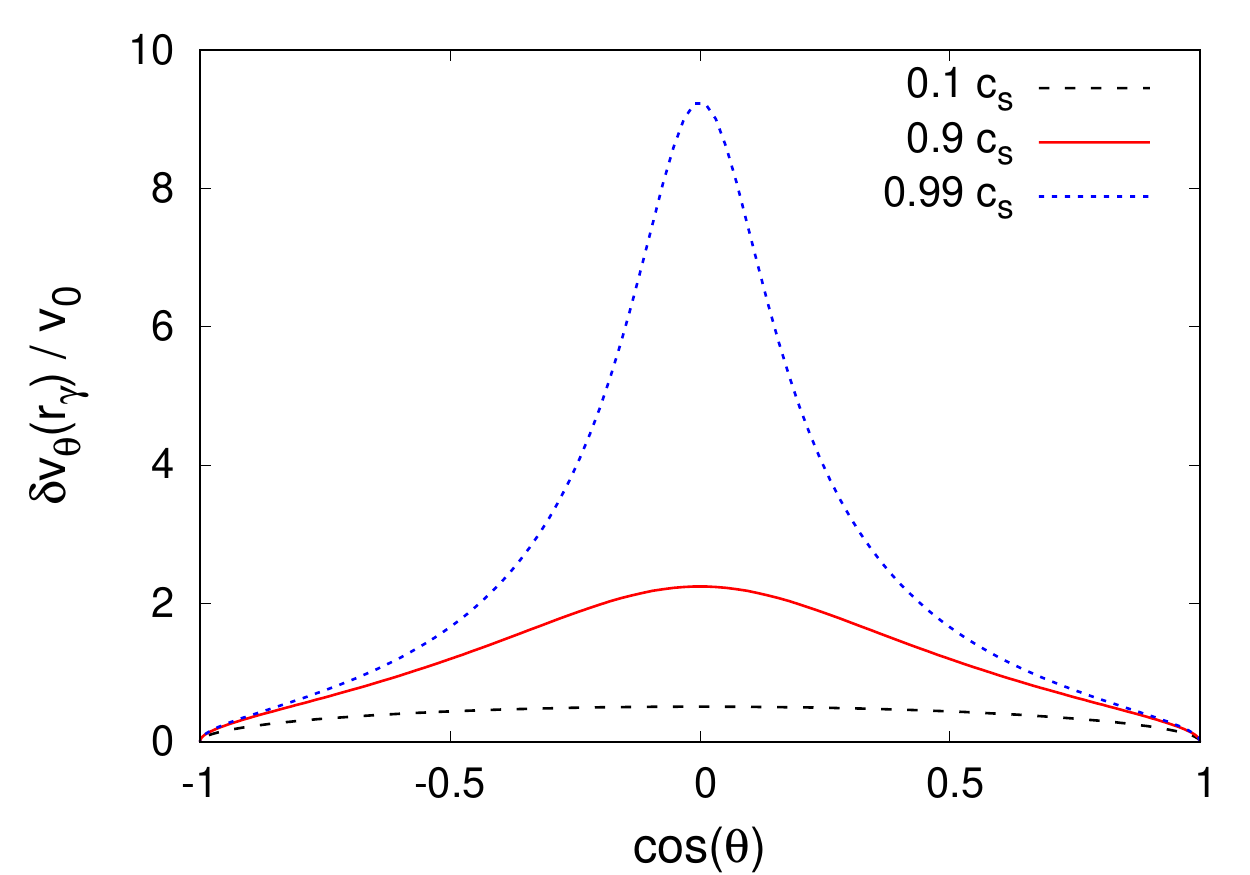}
\end{center}
\caption{
Linear perturbation $\delta v_{\theta}$ to the angular velocity, from Eq.(\ref{eq:delta-v-theta-odd}),
normalized by $v_0$. We show the cases $v_0=0.1 c_s$, $0.9 c_s$ and $0.99 c_s$,
at the radius $\hat r_\gamma$.
}
\label{fig_beta-elliptic}
\end{figure}


We show the angular velocity $\delta v_{\theta}$ in Fig.~\ref{fig_beta-elliptic}.
For $v_0 \to 0$ we have $\delta\hat\beta_{\rm odd} \propto  -\cos \theta$ and
$\delta v_{\theta} \propto (\sin\theta)/r$ and we recover the linear flow
(\ref{eq:beta_L-def}) as multipoles beyond the dipole become negligible, as seen
from (\ref{eq:deltabeta_l_v0}).
By symmetry, the angular velocity vanishes at $\theta=0$ and $\theta=\pi$,
along the $z$-axis.
The height of the central peak grows with $v_0$ and diverges for $v_0 \to c_s$
as $1/\mu$.
The angular velocity is positive. This means that for $\theta \simeq \pi/2$ the
first-order perturbation $\delta\vec v$ is opposite to the incoming flow
$\vec v_0$, which is thus slowed down close to the BH, before turning and falling
increasingly fast into the BH close to the Schwarszchild radius.
We can see that the singularity associated with the transition to the supersonic regime
only appears very close to $c_s$, for $v_0 \gtrsim 0.9 c_s$.

This behavior agrees with that of the linear flow (\ref{eq:beta_L-def}),
which behaved at large radii as
\be
r \gg r_\gamma : \;\;\; \beta_0^L \propto 1/r , \;\;\; \beta_1^L = v_0 r - \frac{v_0}{2\gamma} + \frac{v_0}{2\gamma^2 r} + \dots ,
\label{eq:2F1-expansion}
\ee
where the subleading term $-v_0/(2\gamma)$ in the dipole generates the positive
angular velocity correction $v_0 \sin(\theta)/(2\gamma r)$.
The difference with the linear flow (\ref{eq:beta_L-def}) is that the cubic nonlinearity
in (\ref{eq:flow-k_+-hatbeta}) generates nonzero contributions to all higher-order
multipoles.

\subsubsection{Phase, velocity and density expansions}

As described in the previous section, at large radii the phase $\hat\beta$
can be expanded as
\be
\hat\beta = v_0 r \cos\theta + \delta\hat\beta_{\rm odd}
+ \delta\hat\beta_{\rm even} ,
\ee
where we decompose over odd and even components n $u=\cos\theta$,
with, for $\hat r \gg \hat r_\gamma$,
\ba
&& \delta\hat\beta_{\rm odd} = \hat\delta\beta^{(0)}_{\rm odd}(\theta)
+ \frac{1}{\hat r} \hat\delta\beta^{(1)}_{\rm odd}(\theta)  +  {\cal O}(1/\hat r^2) ,
\nonumber \\
&& \delta\hat\beta_{\rm even} = \frac{1}{\hat r}
\hat\delta\beta^{(1)}_{\rm even}(\theta) + {\cal O}(1/\hat r^2) ,
\label{eq:beta-large-odd-even}
\ea
and $\hat\delta\beta^{(0)}_{\rm odd}$ is given by
Eq.(\ref{eq:deltabeta-odd-large-r-low-v0}) whereas $\hat\delta\beta^{(1)}_{\rm even}/\hat r$
is given by Eq.(\ref{eq:deltabeta-even-large-r}).
As described in App.~\ref{sec:mode-coupling-large-radii}, these large-distance tails,
generated by nonlinear mode couplings, can be expanded in Legendre multipoles,
\ba
&& \delta\hat\beta^{(0)}_{\rm odd}(\theta) = \sum_{\ell=0}^{\infty}
a_{2\ell+1} P_{2\ell+1}(\cos\theta) ,  \nonumber \\
&& \delta\hat\beta^{(1)}_{\rm even}(\theta) =
\sum_{\ell=0}^{\infty} b_{2\ell} P_{2\ell}(\cos\theta) ,
\label{eq:beta-large-r-a-b}
\ea
where the coefficients $a_n$ and $b_n$ obey the recursions (\ref{eq:a_ell_odd_recursion}) and (\ref{eq:b_ell_even_sum}).
The coefficients $a_n$ and $b_n$ remain of the same order as $a_1$ and $b_0$ if
$v_0 \sim c_s$, or decay at high $n$ as $(v_0/c_s)^n$ if $v_0 \ll c_s$.
Thus, for small velocities $v_0 \ll c_s$, we recover the linear flow as higher orders become
negligible and the coefficients $a_1$ and $b_0$ take their linear-flow values.

The velocity field is given by $\vec v = \hat\nabla\hat\beta$, which yields
\ba
v_r & = & v_0 \cos\theta - \frac{1}{\hat r^2} (\delta\hat\beta^{(1)}_{\rm odd}
+ \delta\hat\beta^{(1)}_{\rm even} ) + \dots , \nonumber \\
v_\theta & = & - v_0 \sin\theta + \frac{1}{\hat r}
\frac{d\delta\hat\beta^{(0)}_{\rm odd}}{d\theta} +
\frac{1}{\hat r^2} \left(
\frac{d\delta\hat\beta^{(1)}_{\rm odd}}{d\theta}
+ \frac{d\delta\hat\beta^{(1)}_{\rm even}}{d\theta} \right)
\nonumber \\
&& + \dots
\label{eq:vr-vtheta-expansion}
\ea
Thus, the deviations from the uniform flow $\vec v_0$ decay as $1/\hat r^2$ for
the radial velocity and as $1/\hat r$ for the angular velocity.
Moreover, the angular velocity and the velocity squared are even up to order
$1/\hat r$.
From Eq.(\ref{eq:rho-gamma-v2}), we obtain for the density
\ba
&& \hat\rho_{\rm even} = \hat\rho_0 + \frac{1}{\hat r}
+ \frac{2 v_0\sin\theta}{\hat r}
\frac{d\delta\hat\beta^{(0)}_{\rm odd}}{d\theta} + \dots , \nonumber \\
&& \hat\rho_{\rm odd} = \frac{2 v_0}{\hat r^2} \left[ \cos\theta
\delta\hat\beta^{(1)}_{\rm even} + \sin\theta
\frac{d\delta\hat\beta^{(1)}_{\rm even}}{d\theta} \right] + \dots
\label{eq:rho-even-odd-expansion}
\ea
where $\hat\rho_0 = 3k_0^2/2=\gamma-v_0^2$.
Thus, the density field is even up to order $1/\hat r$.

Using the explicit expression (\ref{eq:deltabeta-odd-large-r-low-v0}) and Eq.(\ref{eq:mu-def-low-v0})
and going back to physical coordinates, we obtain
\ba
&& \rho_{\rm even} = \rho_0 + \frac{{\cal G} M_{\rm BH} \rho_0}
{c_s\sqrt{(c_s^2 - v_0^2) r^2 + v_0^2 z^2}}  + \dots , \nonumber \\
&& \rho_{\rm odd} = \frac{4 B \rho_0 {\cal G}^2 M_{\rm BH}^2 v_0 c_s z}
{ [ (c_s^2 - v_0^2) r^2 + v_0^2 z^2 ]^{3/2} } + \dots
\label{eq:rho-even-odd-explicit}
\ea
The even component agrees with the results of \cite{Dokuchaev1964,Lee:2011px}
for the linear density perturbation in an isothermal gas by a moving star, without mass accretion.
The new odd component, proportional to the coefficient $B$, is related to the accretion
by the BH, as described in Sec.~\ref{sec:accretion} and Eq.(\ref{eq:B-F_star}) below.
It is also the source of the dynamical friction, as shown in Sec.~\ref{sec:friction}.

Expanding $\hat\rho_{\rm even}$ in powers of $v_0$, we have
\be
\hat\rho_{\rm even} = \hat\rho_0 + \frac{1}{\hat r} + \frac{v_0^2}{2 c_s^2 \hat r} \sin^2\theta + \dots
\ee
Thus, at large radii, $\hat r \gg \hat r_\gamma$, and for $v_0 \ll c_s$, the density correction due
to the motion of the BH remains much smaller than the static contribution associated
with the BH, $\hat\rho_0 \gg 1/\hat r \gg v_0^2/(c_s^2 \hat r)$.
Therefore, it is legitimate to neglect this correction to the self-gravity of the dark matter perturbation,
as we assume throughout this paper (note that the $1/\hat r$ term includes the self-gravity
in the response of the scalar cloud to the BH in the static case, see Eq.(\ref{eq:rho-sin-cos})).
At smaller radii, the BH gravity dominates over the scalar-field background
self-gravity, and hence over the scalar perturbation too.

\subsection{Numerical scheme}
\label{sec:nonlinear-flow}

In the subsonic regime that we study in this paper, the flow remains close to the linear
solution (\ref{eq:beta_L-def}). In particular, there is no shock at large radii.
Then, an iterative approach starting from this linear approximation converges and provides
an efficient numerical scheme.
In practice, we write Eq.(\ref{eq:flow-k_+-hatbeta}) as
\be
\hat \nabla \cdot ( k_+^2 \hat\nabla \hat\beta) = S , \;\;\; S = \frac{2}{3} \hat\nabla \cdot
[ (\hat\nabla\hat\beta)^2 \hat\nabla\hat\beta ] ,
\label{eq:system-source}
\ee
and we solve this system (\ref{eq:system-source}) with an iterative scheme.
First, starting from the linear flow $\hat\beta^L$, we compute the source term $S$
from the second equation.
Second, we obtain an improved flow $\hat\beta$ by solving the first equation.
We expand the fields over Legendre multipoles and use the Green's function of the linear operator
$\hat \nabla \cdot [ k_+^2 \hat\nabla (\cdot) ]$, as detailed in App.~\ref{sec:Green}.
We repeat these two steps until the flow converges.

\subsection{Numerical results}
\label{sec:numerical}

For our numerical computation we take $k_0 = 10^{-3}$, which is the order of magnitude
associated with gravity-self-interactions equilibrium.
Indeed, from Eqs.(\ref{eq:hydro-alpha}) and (\ref{eq:k2-infinity}), we have in the bulk of the
soliton, far from the BH horizon, $k^2 \sim \Phi_{\rm I} \sim \Phi_{\rm N}$, and the
typical amplitude of the gravitational potential in astrophysical and galactic systems
is $\Phi_{\rm N} \sim 10^{-6}$.
In virialized systems dominated by gravity, the typical velocities are also of order
$v^2 \sim \Phi_{\rm N}$. This is also the magnitude of the speed of sound, as $c_s^2 \sim k_0^2$.
Because we focus on the subsonic regime in this paper, we take $v_0 = c_s/2$ in the
numerical computations below. This is of the expected order of magnitude while remaining
below the sound speed.
For the matching radius we take $\hat r_{\rm m} = 80$, in order to respect the constraints
discussed in Sec.~\ref{sec:low-high-branches}.
The behavior of the scalar-field flow does not depend on these precise values but only on
the properties $k_0 \ll 1$ and $v_0 < c_s$.

\subsubsection{Legendre multipoles of the velocity field}

\begin{figure}
\begin{center}
\includegraphics[width=\columnwidth]{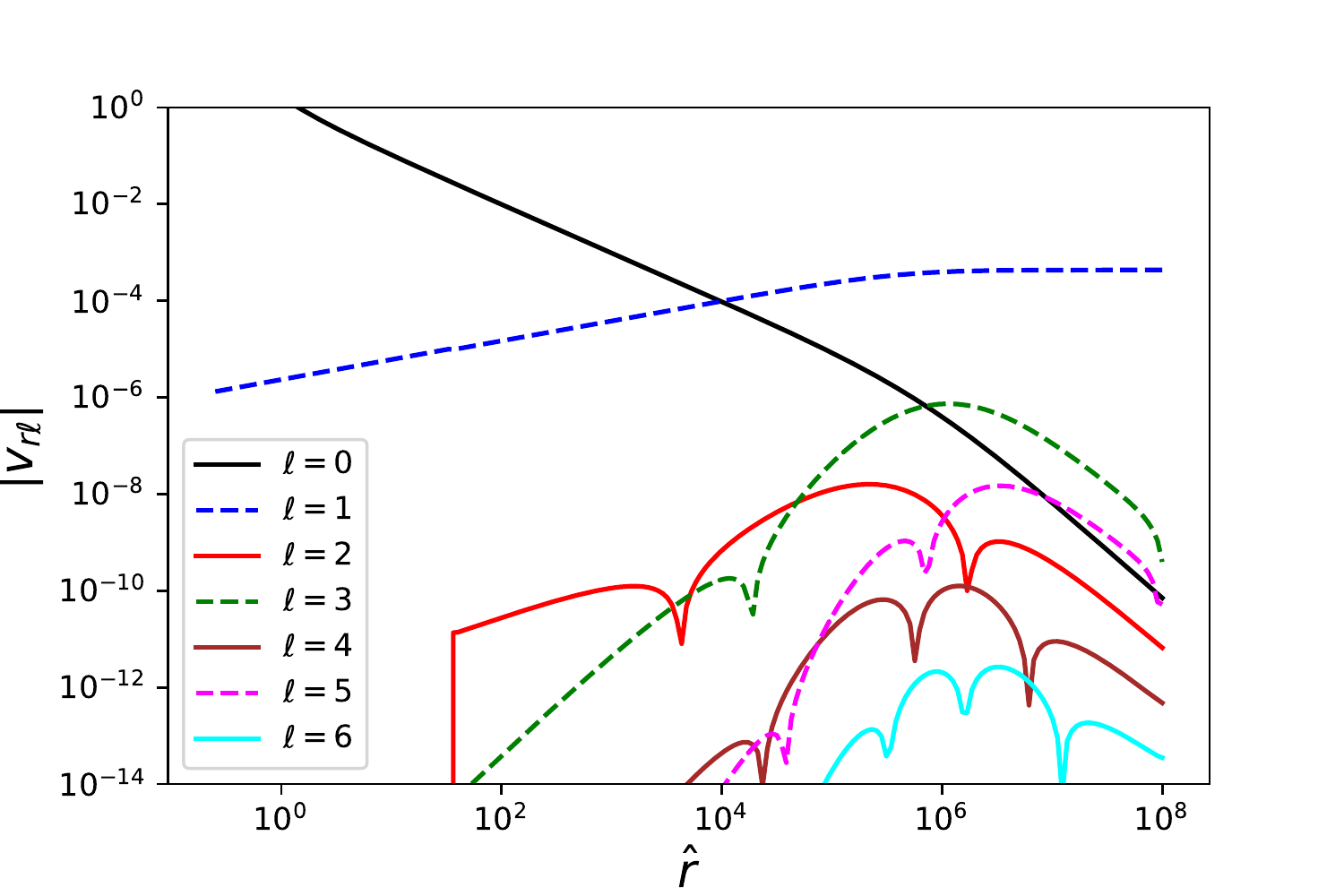}
\includegraphics[width=\columnwidth]{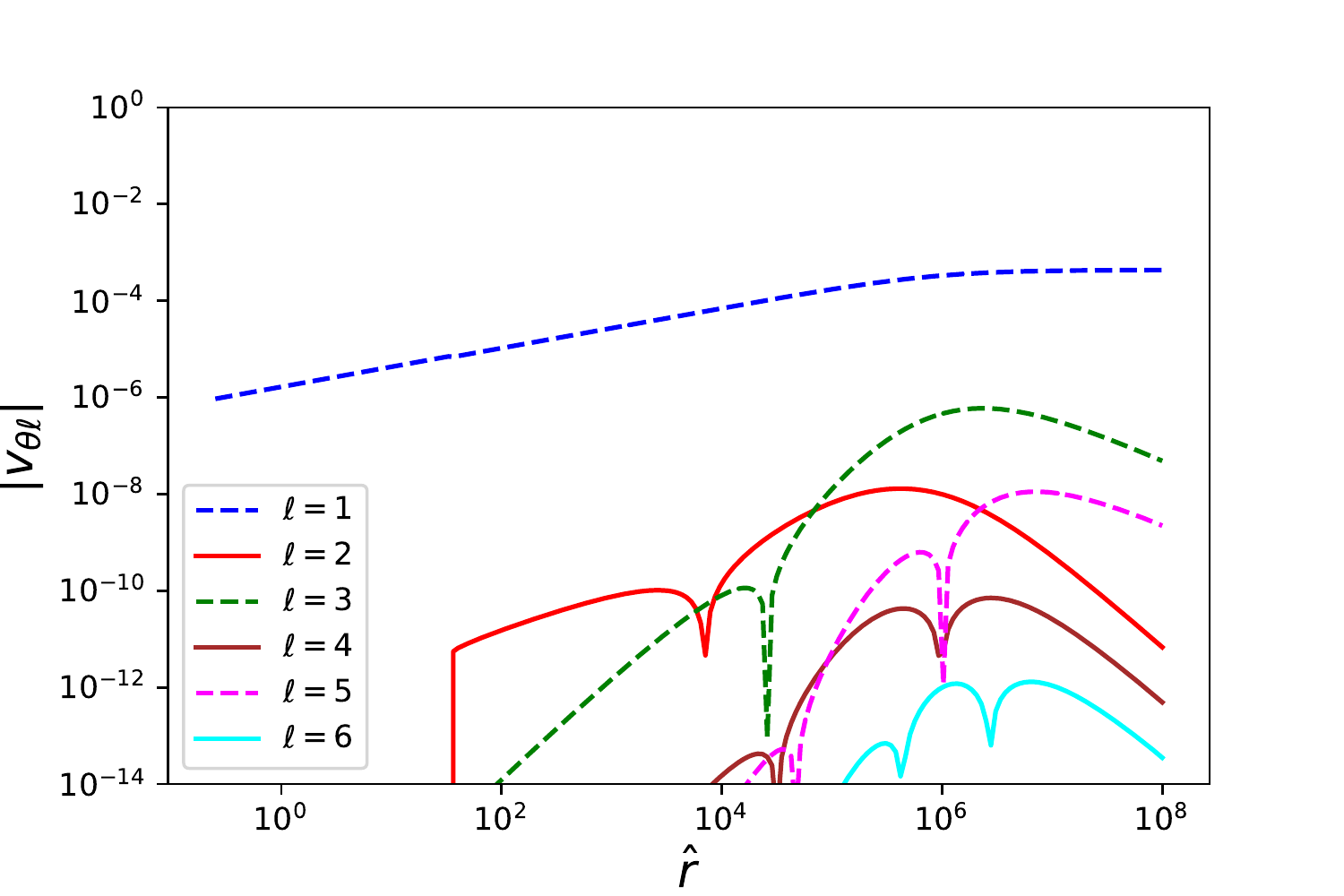}
\end{center}
\caption{
First Legendre multipoles of the radial velocity $v_{r\ell}$ (upper panel) and of the angular velocity
$v_{\theta\ell}$ (lower panel), as defined in Eq.(\ref{eq:vr-vtheta-ell-def}).
The solid lines show even indices $\ell$ whereas the dashed lines show
odd $\ell$.
}
\label{fig_dbeta_dr}
\end{figure}

We obtain the Legendre multipoles of the phase $\beta$ by the iteration scheme described
previously, below Eq.(\ref{eq:system-source}). More details are provided in App.~\ref{sec:Green}.
We find that the scheme converges after a few iterations and we show our results obtained
after the 9th iteration. This is because the nonlinear flow remains close to the linear approximation
(\ref{eq:beta_L-def}) in the subsonic regime, $v_0 < c_s$.

We show in Fig.~\ref{fig_dbeta_dr} our results for the Legendre multipoles of the radial
and angular velocities, defined as
\be
v_r = \sum_{\ell} v_{r\ell} P_\ell(\cos\theta) , \;\;\;
v_{\theta} = -\sin\theta \sum_{\ell} v_{\theta\ell} P'_\ell(\cos\theta) ,
\label{eq:vr-vtheta-ell-Pell}
\ee
with
\be
v_{r\ell} = \frac{d\hat\beta_\ell}{d\hat r} , \;\;\; v_{\theta\ell} = \frac{\hat\beta_\ell}{\hat r} .
\label{eq:vr-vtheta-ell-def}
\ee

At small radii, $r<r_{\rm m}$, we use the radial monopole obtained from radial accretion
in \cite{Brax:2019npi} and the dipole obtained for the linear flow (\ref{eq:beta_L-def}).
This sets the inner boundary condition and we only solve the nonlinear system
(\ref{eq:system-source}) for $r > r_{\rm m}$.
This is why the higher-order multipoles are truncated at $\hat r_{\rm m}$.
We can check that this procedure is valid as they are indeed negligible as compared with
the monopole at radius $\hat r_{\rm m}$.
The radial velocity $v_r$ diverges at the Schwarzschild radius \cite{Brax:2019npi},
but this is an artefact due to the choice of coordinates and to the fact that the quantity
$d\hat\beta/d\hat r$ can only be interpreted as a velocity in the nonrelativistic regime.

We can see in the figure the partial decoupling of odd and even components described
in App.~\ref{sec:mode-coupling-large-radii} and Sec.~\ref{sec:Mach-large-radii}.
Odd multipoles of the phase $\hat\beta$ have a constant
tail at large distance, in addition to the linear dipole associated with the uniform flow $\vec v_0$,
whereas even multipoles decrease as $1/\hat r$, see
Eqs.(\ref{eq:beta-large-odd-even})-(\ref{eq:beta-large-r-a-b}).
This implies that for the angular velocity, $v_\theta=(1/\hat r) \partial\hat\beta/\partial\theta$,
the even multipoles decay as $1/\hat r$ whereas the odd multipoles decay as $1/\hat r^2$,
see Eq.(\ref{eq:vr-vtheta-expansion}).
Note that with the notation (\ref{eq:vr-vtheta-ell-Pell}) even components of $v_\theta$
correspond to odd $\ell$.
These two different decay rates are clearly seen in the lower panel, where
solid lines show the odd components and dashed lines the even components.

On the other hand, the constant odd tail of the phase $\hat\beta$ does not contribute
to the radial velocity, $v_r=\partial\hat\beta/\partial\hat r$.
This implies that only the leading even tail and the subleading odd tail of order $1/\hat r$ in
$\hat\beta$ contribute, and all multipoles of the radial velocity decay at the same
rate $1/\hat r^2$, as seen in the upper panel and in Eq.(\ref{eq:vr-vtheta-expansion}).

We can see that the large-distance asymptotic regime is only reached beyond
$\hat r_\gamma \sim 10^6$, and at larger radii for higher-order multipoles.
This radius $r_\gamma \sim r_{\rm sg}$ corresponds to the radius where the scalar-field
self-gravity and pressure are of the same order as the BH gravity.
At larger radii, the impact of the BH gravity is screened by the collective response of the
scalar field (its pressure). This leads to the steep falloff of the velocity
corrections to the uniform flow $\vec v_0$. We will see below that the same physics
also damps the perturbations of the density field at large radii and replaces the divergent
Coulomb logarithm of the collisionless dynamical friction by a finite number (while also
changing the scaling in $v_0$ and $c_s$).

We also note that even for the relatively large velocity $v_0 = c_s/2$, the corrections
to the linear flow (\ref{eq:beta_L-def}) are still small, as shown by the magnitude of the
higher-order multipoles $\ell \geq 2$.
Thus, as will also be apparent in Fig.~\ref{fig_density} below, the flow appears as a superposition
of a monopole radial accretion, close to the purely radial result, with a dipole term associated
with the uniform velocity $\vec v_0$ at infinity.
This agrees with the numerical results obtained in \cite{Hunt1971,Petrich1989} for the motion
of a BH in a perfect gas, with either Newtonian or relativistic treatments.

\subsubsection{Legendre multipoles of the density field}

\begin{figure}
\begin{center}
\includegraphics[width=\columnwidth]{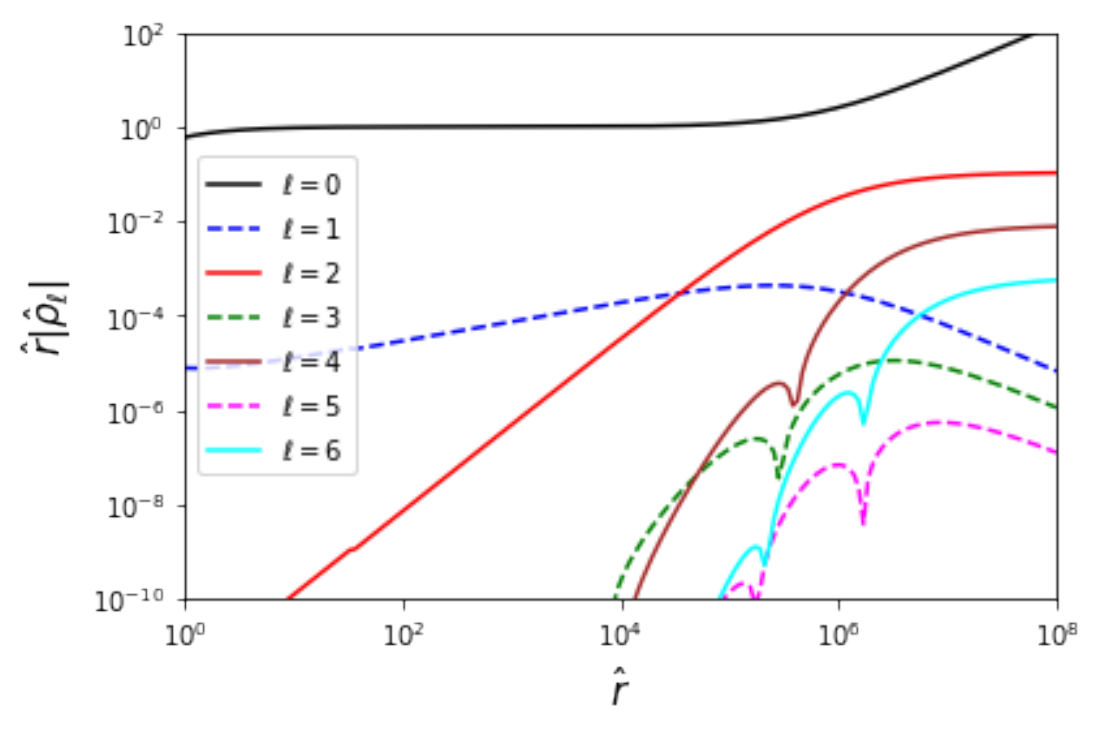}
\end{center}
\caption{
First Legendre multipoles of the density multiplied by a factor $\hat r$, $\hat r \hat\rho_{r\ell}$.
}
\label{fig_rho-ell}
\end{figure}

We show in Fig.~\ref{fig_rho-ell} our results for the Legendre multipoles
$\hat\rho_\ell$ of the density field, multiplied by a factor $\hat r$.
We can see that the monopole dominates at all radii. This is because
the density goes to the constant density $\hat\rho_0$ associated with the
soliton at large radii, whereas at small radii the flow becomes radial,
as seen in Fig.~\ref{fig_dbeta_dr}, which also implies a spherically-symmetric
configuration.

In agreement with Eq.(\ref{eq:rho-even-odd-expansion}), the product
$\hat r \hat\rho_{r\ell}$ goes to a constant at large radii for the even multipoles beyond $\ell=0$,
and decreases as $1/\hat r$ for the odd multipoles.
At small $\hat r$, the monopole grows as $1/\hat r$ while higher-order multipoles
grow more slowly or decrease.
In particular, for the linear flow (\ref{eq:beta_L-def}), using
Eq.(\ref {eq:beta0-beta1-small-r}) we can see that the dipole behaves as
$\hat r^{\sqrt{2}-2}$ at small radii, so that $\hat r \hat\rho_{\ell=1} \propto \hat r^{\sqrt{2}-1} \to 0$.

\subsubsection{Scalar-field flow and density maps}
\label{sec:maps}

\begin{figure*}
\begin{center}
\includegraphics[width=\textwidth]{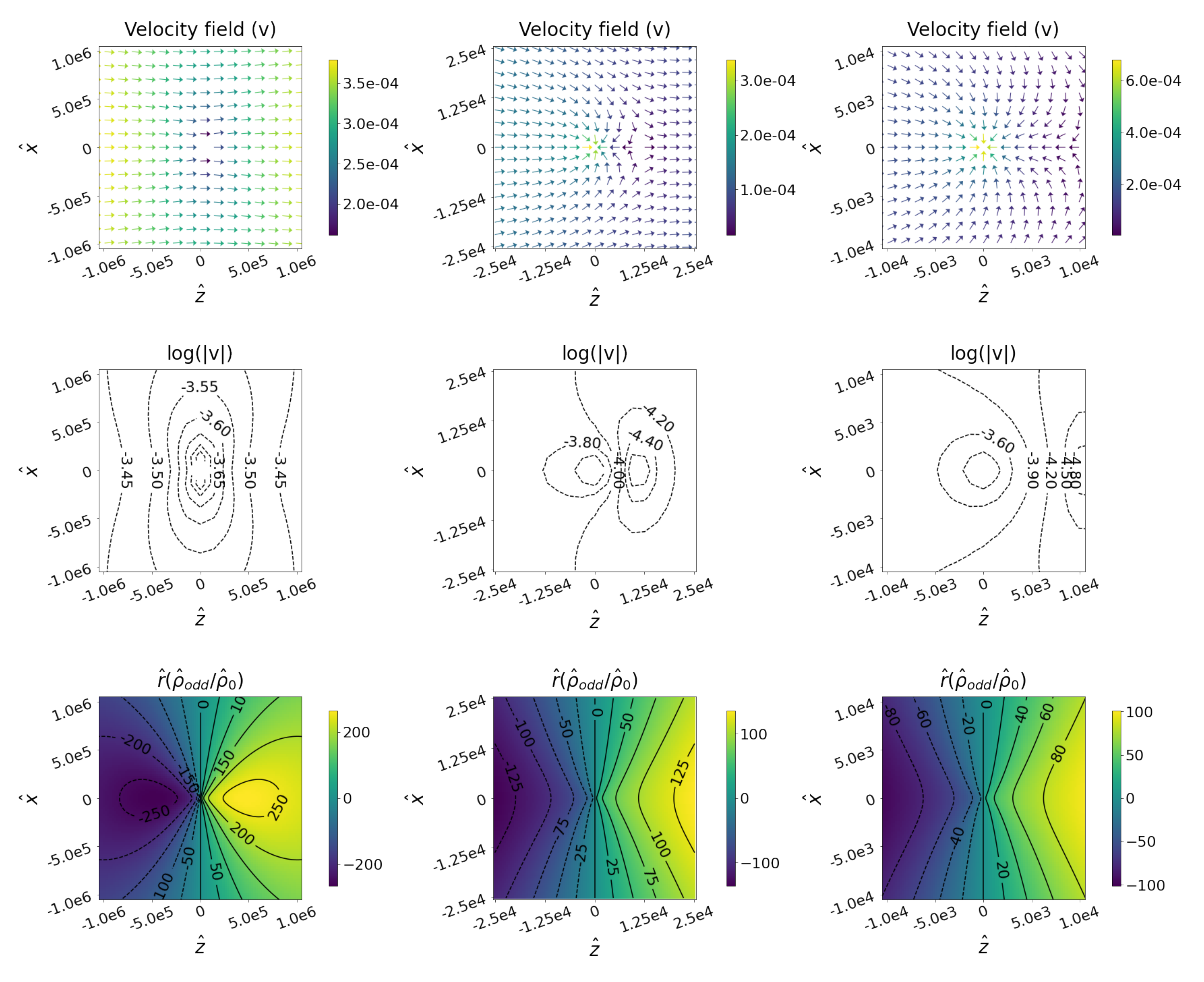}
\end{center}
\caption{Flow (top panels), iso-velocities contours (middle panels) and odd-component of the
density field $\hat r \hat\rho_{\rm odd}/\hat\rho_{0}$ (bottom panels) for the scalar-field at different
scales ($10^6$, $2.5 \times 10^4$ and $10^4$  $r_s$). The velocity and
the density are computed from the multipoles of $\hat{\beta}$.
The BH is located at the center of the figures, at $\hat{z}=\hat{x}=0$,
where $\hat{z}=z/r_{\rm s}$ and $\hat{x}=x/r_{\rm s}$.
}
\label{fig_density}
\end{figure*}

We display in Fig.~\ref{fig_density} maps of the scalar velocity and density fields.
The BH is located at the center of the plots and the scalar-field dark matter comes from the
left with the uniform velocity $\vec v_0$ and density $\rho_0$.

The upper row shows maps of the velocity field, represented by the arrows with unit length,
on different scales. We zoom in onto the BH from the left to the right panel.
The middle row shows the corresponding maps of the magnitude $|\vec v|$ of the velocity.

On the largest scale, in the left panels, the flow is almost unaffected by the gravitational
pull from the BH and roughly keeps its incoming velocity $\vec v_0$, moving from the left to the right.
As seen in the middle-row panel below, the flow is almost symmetric over
$\hat z \leftrightarrow - \hat z$, that is, $v^2$ is an almost even function of $\cos\theta$.
This agrees with the large-radius expansions analysed in Sec.~\ref{sec:Mach-large-radii}
and App.~\ref{sec:mode-coupling-large-radii} and shown in Fig.~\ref{fig_dbeta_dr}.
From Eq.(\ref{eq:vr-vtheta-expansion}), at large distance the first correction
$\delta\vec v$ to the velocity, with respect to the incoming
velocity $\vec v_0$, is an even angular velocity
$\delta v_{\theta {\rm even}} \propto 1/\hat r$.
This gives an even correction to the velocity magnitude,
$\delta v^2_{\rm even} \propto 1/\hat r$.
This agrees with the pattern seen in the left panels. The streamlines slightly focus towards
the BH in a symmetric fashion with respect to $\hat z \leftrightarrow - \hat z$.
This is obviously different from the case of free point particles and similar to what happens
in a gas. This is due to the pressure generated by the self-interactions, that grows near the
BH as the density increases. At large distance, the main effect is that particles are first slowed
down, as they approach $\hat z=0$, and next accelerated to recover the velocity $\vec v_0$
downstreams.

In the middle-column panels, on intermediate scales, we can see more clearly the streamlines
being deflected towards the BH. We can also see a turning point on the $\hat z$-axis somewhat
behind the BH. This separates the region, far from the BH, where the streamlines escape
to infinity to the right of the figure, and the inner region where the streamlines fall into the BH.
Obviously, there is no such turning point to the left of the BH, as the dark matter coming
from the left along the $\hat z$-axis keeps moving straight towards the BH until it falls into the latter.
This is a clear signature of the asymmetry of the flow, in contrast with the case of a potential
flow around a compact object without accretion (e.g. the flow of water around a hard ball).
Mathematically, this is due to the different boundary conditions around the object,
the radial infall at the BH horizon in our case or the vanishing normal velocity at the surface
of the ball in the usual hydrodynamical case.
As seen in previous sections, the boundary condition at infinity corresponds to the dipole
in Eq.(\ref{eq:beta_L-def}), while the boundary condition close to the center corresponds
to the monopole in Eq.(\ref{eq:beta_L-def}). Thus, the two boundary conditions have different
parity, which implies the flow is neither exactly odd or even. The phase is odd at large
distance and becomes even close to the BH, with a complex pattern in the intermediate region.
This also means that the asymmetry of the flow is related to the accretion by the BH, which
determines the inner boundary condition.
The dynamical friction of the BH, due to this asymmetry, is therefore directly related to the
accretion rate. We will recover this relationship in Sec.~\ref{sec:friction}, where we obtain
the explicit expression of the dynamical friction.

In the right-column panels, we can see the flow becoming radial as we zoom in closer to the BH.
This agrees with the results of Fig.~\ref{fig_dbeta_dr}, which show that the monopole dominates
at small radii. The velocity magnitude grows at smaller radii as the flow is accelerated by the
BH gravity during its infall. As explained in Sec.~\ref{sec:low-high-branches},
below a critical radius $r_c$ the flow switches to the high-velocity branch, the pressure
due to the self-interactions is no longer able to resist gravity and the dark matter reaches the
BH horizon as in free fall.

The lower row in Fig.~\ref{fig_density} shows maps of the odd component of the density field,
more precisely the ratio $\hat r\hat \rho_{\rm odd}/\hat\rho_{0}$.
We single out the odd component to emphasize the asymmetry in the flow and the appearance
of a wake behind the BH. Indeed, the dynamical friction of the BH is due to the asymmetry of the
flow (by symmetry, a symmetric flow would not generate any drag force) but it would be
difficult to distinguish it in a map of the total density, as the even component dominates on all scales as was found in Fig.~\ref{fig_rho-ell}.
Indeed, the total density appears almost spherically symmetric on all scales in the subsonic
regime that we consider in this paper. The same pattern is found for the case of a perfect
gas, in both Newtonian and relativistic numerical simulations \citep{Hunt1971,Petrich1989}.

We add the factor $\hat r$ to see more clearly the radii that dominate the gravitational pull on the BH
by the overdense wake of the dark matter, as the gravitational force is of the form
$F_{\rm grav} \sim \int d\vec r \rho \vec r/r^3 \sim \int d\ln r (r \rho)$.
From Eq.(\ref{eq:continuity-Bernouilli}) a decrease of the velocity implies an increase of the density,
as $\hat \rho = 3 k_+^2(\hat r)/2 - v^2$.
Then, the turning point in the velocity field, somewhat behind the BH, corresponds to
an increase of the density as compared with the radial reference.
This clearly shows the asymmetry and the existence of a wake behind the BH.
As seen in the figure, and in agreement with Fig.~\ref{fig_rho-ell},
the product $\hat r \hat\rho_{\rm odd}$ peaks at the large radius $\hat r_\gamma \sim 10^6$.
This is also the radius $\hat r_{\rm sg}$ where the soliton self-gravity becomes of
the order of the BH gravity.
Thus, in contrast with the case of free particles, beyond $\hat r_\gamma$ the self-interaction
pressure dominates over the BH gravity and screens its impact on the dark matter distribution.
This provides a large-scale cutoff, which will also remove the Coulomb logarithm of the dynamical
friction found for collisionless particles in Chandrasekhar's classical study
\citep{Chandrasekhar:1943ys}.

\section{ Mass accretion by the BH}
\label{sec:accretion}

\subsection{Relationship with large-radius expansions}

\subsubsection{Mass flow through a large sphere}

In a steady state, the accretion of matter by the BH is given by the flux of matter through any closed
surface that surrounds the BH,
\be
\dot {\hat M}_{\rm BH} = - \int_{\hat S} \vec{d \hat{S}} \cdot \hat\rho \vec {v} ,
\label{eq:dot-M_BH-S}
\ee
where the radius of the surface is large enough for the low-$k$ nonrelativistic regime
(\ref{eq:continuity-Bernouilli}) to apply.
We can check that the accretion rate $\dot{\hat M}_{\rm BH}$ does
not depend on the surface $\hat S$, as the difference between fluxes through $\hat S$ and
any smaller or greater surface $\hat S'$ is given by the integral of $\hat\nabla\cdot(\hat\rho \vec v)$
over the volume $\hat V_{\hat S,\hat S'}$ between these two surfaces, which is zero from Eq.(\ref{eq:continuity-Bernouilli}).
This means that we can obtain the mass loss of the scalar field medium from the large-distance
expansion (\ref{eq:beta-large-odd-even}), by choosing a surface much beyond the radius
$\hat r_\gamma$.
Choosing for $\hat S$ the sphere of radius $\hat r$, the mass flux reads
\be
\dot {\hat M}_{\rm BH} = - 2 \pi \hat r^2 \int_{-1}^1 du \, \hat\rho v_r ,
\label{eq:dot-M-sphere-int}
\ee
where $u=\cos\theta$. Thus, only the monopole of the radial momentum $\hat\rho v_r$
contributes. At leading order $1/\hat r^2$, it is set by the even $b_{2\ell}$ series in
(\ref{eq:beta-large-r-a-b}) and we obtain
\be
\dot {\hat M}_{\rm BH} = 4 \pi \left[ b_0 \left( \gamma - \frac{5}{3} v_0^2 \right)
+ b_2 \frac{8 v_0^2}{15} \right] ,
\label{eq:dot-M_BH-sphere}
\ee
which is independent of the radius $\hat r$ as it should.

\subsubsection{Mass flow following the streamlines}

\begin{figure}
\begin{center}
\includegraphics[width=\columnwidth]{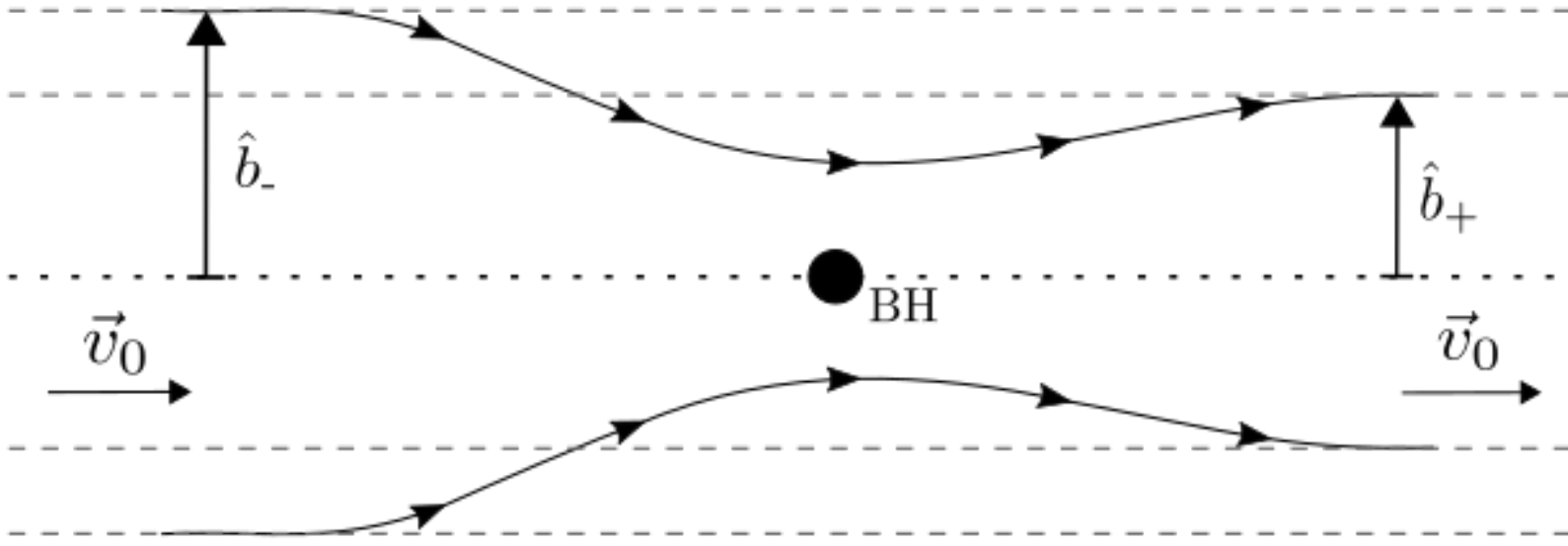}
\end{center}
\caption{
The axisymmetric cylinder following the streamlines used for the computation
of Eq.(\ref{eq:dot-M-BH-cyl}).
}
\label{fig_vol-accretion}
\end{figure}

Alternatively, we can compute the accretion of mass by the BH by following the scalar-field
streamlines, in a manner similar to the computation in the case of free particles where one follows
their trajectories.
In contrast with the free case, the streamlines do not escape to $\hat z \to \infty$ with
a nonzero deflection angle $\theta_\infty$. Thanks to  the effective pressure generated by the
self-interactions, the flow is smooth, without shocks or caustics, as long as we restrict ourselves to the
subsonic regime (\ref{eq:subsonic}), and the velocity at large distances downstream is again
$\vec v_0$. Thus, the streamlines are again parallel to the $z-$axis far downstream,
see the velocity field shown in the upper row in Fig.~\ref{fig_density}.
Then, as shown in Fig.~\ref{fig_vol-accretion},
we take for the closed surface $\hat S$ the cylinder of varying  transverse radius
$\hat b(\hat z)$ that follows the streamlines of incoming impact parameter $\hat b_-$ at
$\hat z_- \to -\infty$ and outgoing transverse radius $\hat b_+$ at $\hat z_+ \to +\infty$.
As discussed in Sec.~\ref{sec:maps}, because of the asymmetry generated by the mass loss
into the BH, the streamlines are not exactly even and $\hat b_+ \neq \hat b_-$.
Then, the mass loss reads
\be
\dot {\hat M}_{\rm BH} = 2\pi \int_0^{\hat b_-} d\hat b \, \hat b \left. \hat \rho v_z \right|_{\hat z_-}
- 2\pi \int_0^{\hat b_+} d\hat b \, \hat b \left. \hat \rho v_z \right|_{\hat z_+} .
\label{eq:dot-M-BH-cyl}
\ee
This is the difference between the upstream and downstream mass fluxes, as there is no mass
flux through the transverse surface of the cylinder.

The streamlines $\hat r(\theta)$ can be obtained by integrating
\be
\mbox{streamlines} : \;\;\; \frac{d\hat r}{d\theta} = \frac{\hat r v_r}{v_\theta} ,
\ee
or from the stream function $\Psi$ defined by
\be
\frac{\partial\Psi}{\partial\theta} = \rho r^2 \sin\theta \, v_r , \;\;\;
\frac{\partial\Psi}{\partial r} = - \rho r \sin\theta \, v_\theta ,
\ee
which ensures the continuity equation (\ref{eq:continuity-Bernouilli}) is satisfied,
$\nabla\cdot(\rho\vec v) = 0$, as the streamlines correspond to the curves
of constant $\Psi$.
From the expansions (\ref{eq:beta-large-r-a-b}) we can obtain the large-distance
expansions of the velocity and density fields, of the streamlines and the stream function.
We obtain
\be
\hat r(\theta) = \frac{\hat b}{\sin\theta} - \frac{1}{\hat b v_0 \sin\theta} \sum_\ell b_{2\ell}
( 1+\cos\theta P_{2\ell} ) + \dots
\ee
where we only kept the leading orders in the impact parameter $\hat b$ of the even and
odd components of the streamlines.
The first term gives at leading order $\hat r\sin\theta=\hat b$. This is the even straight line of
constant transverse radius $\hat b$, parallel to the $z-$axis, associated with the zeroth-order
solution when the BH gravity is neglected.
The second term is the first asymmetric contribution, which is a subleading correction as there
is also an even term of order $\hat b^0$ generated by the $a_{2\ell+1}$ series in the expansion
(\ref{eq:beta-large-r-a-b})
(as noticed in Sec.~\ref{sec:Mach-large-radii} and Eq.(\ref{eq:vr-vtheta-expansion}), the leading
correction to the flow is even in $\hat z \leftrightarrow - \hat z$ and the asymmetry only appears
at the next subleading order).
The impact parameter upstream is $(\hat r\sin\theta)_{\hat z\to -\infty}= \hat b$
while the transverse radius downstream is
\be
(\hat r\sin\theta)_{\hat z\to +\infty}= \hat b -  \frac{2}{\hat b v_0} \sum_\ell b_{2\ell} + {\cal O}(1/\hat b^2) .
\label{eq:b-downstream}
\ee
The series in $b_{2\ell}$ can be expressed in terms of $b_0$ and $b_2$ by using
the explicit expression (\ref{eq:deltabeta-even-large-r}) and noticing that
$\delta\hat\beta^{(1)}_{\rm even}(\theta=0) = B = \sum_\ell b_{2\ell}$.
This gives
\be
B = \sum_\ell b_{2\ell} = \frac{b_0}{3} \frac{3\gamma-5 v_0^2}{\gamma-v_0^2} + \frac{4 b_2}{15}
\frac{2 v_0^2}{\gamma-v_0^2} .
\label{eq:series-b-2ell}
\ee
In fact, using the explicit expression (\ref{eq:deltabeta-even-large-r}) we can also express $b_0$
and $b_2$ in terms of $B$. This gives in particular
\be
b_0 = B \frac{c_s}{2 v_0} \ln \left( \frac{c_s+v_0}{c_s-v_0} \right)
= B \left[ 1 + \frac{v_0^2}{3 c_s^2} + \dots \right] .
\label{eq:b0-B}
\ee

Going back to the expression (\ref {eq:dot-M-BH-cyl}) for the accretion rate,
we can choose the large-distance limit of the cylinder such that
$|\hat z_{\pm}| \gg \hat b_{\pm} \gg \hat r_\gamma$.
Then, the angles $\theta_-$ and
$\theta_+$ go to $\pi$ and $0$, the density and velocity go to
$\hat \rho_0 = \gamma - v_0^2$ and $\vec v_0$, and we obtain
\be
\dot {\hat M}_{\rm BH} = \pi \hat\rho_0 v_0 ( \hat b_-^2 - \hat b_+^2 ) = 4 \pi \hat \rho_0
\sum_\ell b_{2\ell} = 4 \pi \hat \rho_0 B ,
\label{eq:dot-M-BH-B}
\ee
where we used Eq.(\ref{eq:b-downstream}) for the relationship between the upstream
and downstream transverse radii $\hat b_{\pm}$
and the first equality in Eq.(\ref{eq:series-b-2ell}).
Using the second equality in Eq.(\ref{eq:series-b-2ell}) we recover
Eq.(\ref{eq:dot-M_BH-sphere}), as it should.

The expression (\ref{eq:dot-M-BH-B}) fully determines the large-distance even component
of the phase $\hat\beta_{\rm even}$ in terms of the BH accretion rate, with
Eq.(\ref{eq:deltabeta-even-large-r}) and $B=\dot{\hat M}_{\rm BH}/(4\pi\hat\rho_0)$.
The latter is set by the boundary conditions close to the BH, at the matching radius $r_{\rm m}$.
As explained in Sec.~\ref{sec:low-high-branches}, at small radii the flow is in the relativistic regime
with a radial  pattern, as the monopole radial velocity grows like $1/r$ whereas higher multipoles
decrease.
At the critical radius $r_c$ the flow makes a smooth transition from the low-velocity branch $v \ll k_+$
to the high-velocity branch $v \simeq k_+$. This also sets the critical value $F_c$ of the
scalar-field flux, which is self-regulated by the pressure associated with the scalar self-interactions.
This gives the connection between the large-distance behavior (\ref{eq:beta-large-odd-even})
and the small-scale relativistic physics near the BH horizon.
Going back to physical coordinates, the flux obtained in this fashion for the radial case reads
\cite{Brax:2019npi}
\be
\dot M_{\rm BH} = 4\pi F_\star \frac{r_s^2 m^4}{\lambda_4} = 3\pi F_\star \rho_a r_s^2 ,
\label{eq:dot-M-BH-radial}
\ee
where the numerical value $F_\star \simeq 0.66$ is obtained from the numerical computation
of the unique profile that goes from the Schwarzschild radius to the outer static soliton.

The comparison of Eqs.(\ref{eq:dot-M-BH-B}) and (\ref{eq:dot-M-BH-radial}) gives
\be
\dot M_{\rm BH} = 4\pi \rho_0 r_s^2 B \;\;\; \mbox{with} \;\;\;
B = F_\star \frac{3\rho_a}{4\rho_0} = \frac{F_\star}{k_0^2} .
\label{eq:B-F_star}
\ee
From Eq.(\ref{eq:beta-large-r-a-b}), the monopole of the radial velocity at large distance
reads $v_{r_0} = - b_0 r_s^2/r^2$. Using (\ref{eq:B-F_star}) and (\ref{eq:b0-B}),
this agrees in the limit $v_0 \to 0$ with the result
$v_r = - F_\star m^4 r_s^2/(\lambda_4 \rho_0 r^2)$ obtained in \cite{Brax:2019npi} for the radial case.

In this paper we focus on the subsonic case, $v_0 < c_s \ll 1$, hence we always have $v_0 \ll 1$.
Then, as explained above, the flow becomes radial much before reaching the critical radius $r_c$
and the self-regulated critical flux $F_c$ is identical to the one obtained in the purely radial case.
This sets the accretion rate by the BH to (\ref{eq:dot-M-BH-radial}), which does not depend on
$v_0$. However, the scalar-field flow at large radii depends on $v_0$, including its monopole
component as seen from Eq.(\ref{eq:b0-B}), with a singularity at $v_0 \to c_s$.
This singularity is beyond the treatment that we give here as we only focus on the
regime $v_0 < c_s$.
As seen in Figs.~\ref{fig_beta-elliptic} and \ref{fig_dbeta_dr}, even at $v_0 = c_s/2$ the flow
remains close to the linear flow (\ref{eq:beta_L-def}), with only small nonlinear corrections.
We checked that our numerical profile gives at large distance a coefficient $b_0$
for the monopole of $\hat\beta$ that agrees with the prediction (\ref{eq:b0-B})-(\ref{eq:B-F_star}).

\subsection{Comparison with previous works and other systems}

The result (\ref{eq:dot-M-BH-radial}) implies
\be
\dot M_{\rm BH} \sim \rho_0 r_s^2/c_s^2 \sim \rho_0 {\cal G}^2 M_{\rm BH}^2 / c_s^2 .
\label{eq:MBdot-approx}
\ee
This is different from the radial accretion of collisionless particles with an isotropic and
monoenergetic distribution at the characteristic velocity $c_s$ \cite{Shapiro:1983du}
\be
\mbox{collisionless:} \;\;  \dot M_{\rm free} =
\frac{16 \pi \rho_0 {\cal G}^2 M_{\rm BH}^2}{c_s}
\label{eq:accretion-free}
\ee
and the classical radial Bondi accretion rate \cite{Bondi:1952ni} for an
isothermal gas, $\dot M_{\rm Bondi} \sim \rho_0 r_s^2/c_s^3$, which also corresponds to
the subsonic limit of the so-called ``Bondi-Hoyle-Lyttleton accretion rate''
\cite{Hoyle1939,Edgar:2004mk}
\be
\mbox{Bondi-Hoyle:} \;\; \dot M_{\rm Bondi-Hoyle} =
\frac{2\pi \rho_0 {\cal G}^2 M_{\rm BH}^2}{(c_s^2+v_0^2)^{3/2}}
\label{eq:accretion-Bondi}
\ee
The hydrodynamical accretion rate (\ref{eq:accretion-Bondi}) is much greater than the
collisionless accretion rate (\ref{eq:accretion-free}), by a factor $(c/c_s)^2 \sim 10^6$,
where $c=1$ is the speed of light. This is because the collisions restrict tangential motion and
funnel particles in the radial direction \citep{Shapiro:1983du}.
The scalar-field accretion rate is in-between these two cases.
As could be expected, for the same hydrodynamical reason it is higher than the free rate,
as the flow is more efficiently converted into a radial pattern at small radii,
but now by a factor $c/c_s \gg 1$.
However, it is much smaller than the accretion rate of the perfect gas rate, by a factor $c_s/c \ll 1$.
Thus, the scalar-field self-interactions are much more efficient
to resist the BH gravity and slow down the infall.
This is because the scalar field has a different equation of state and deviates from a perfect gas
in the relativistic regime, which sets the critical flux $F_c$ and the normalization of the global profile
\citep{Brax:2019npi}.
This agrees with the fact that for a perfect gas with adiabatic index $\gamma_{\rm ad} > 5/3$,
there is no Newtonian steady transonic solution but one exists in General Relativity
\citep{Shapiro:1983du,Petrich1989}.
This again shows the critical role of relativistic effects at small radii for steep equations of state.

The expression (\ref{eq:dot-M-BH-radial}) can be understood in simple terms.
It simply means that close to the BH horizon $r_s$,
where the infall velocity is close to the speed of light, the scalar density is of the order of $\rho_a$,
as can be checked by an explicit computation of the scalar profile, see \cite{Brax:2019npi}
and Eq.(\ref{eq:radial-profile-rho-vr}).
From Eq.(\ref{eq:Phi_I-psi}), this is the density where the self-interaction potential $\Phi_{\rm I}$
is of order unity and the self-interaction term $V_{\rm I} = \lambda^4\phi^4/4$ is of the order
of the mass term $m^2\phi^2/2$. This characteristic density provides an upper bound
on $\rho$, and hence on the accretion rate, as the infall velocity cannot be greater than the
speed of light.

\section{Dynamical friction}
\label{sec:friction}

\subsection{Relationship with large-radius expansions}

\begin{figure}
\begin{center}
\includegraphics[width=0.8\columnwidth]{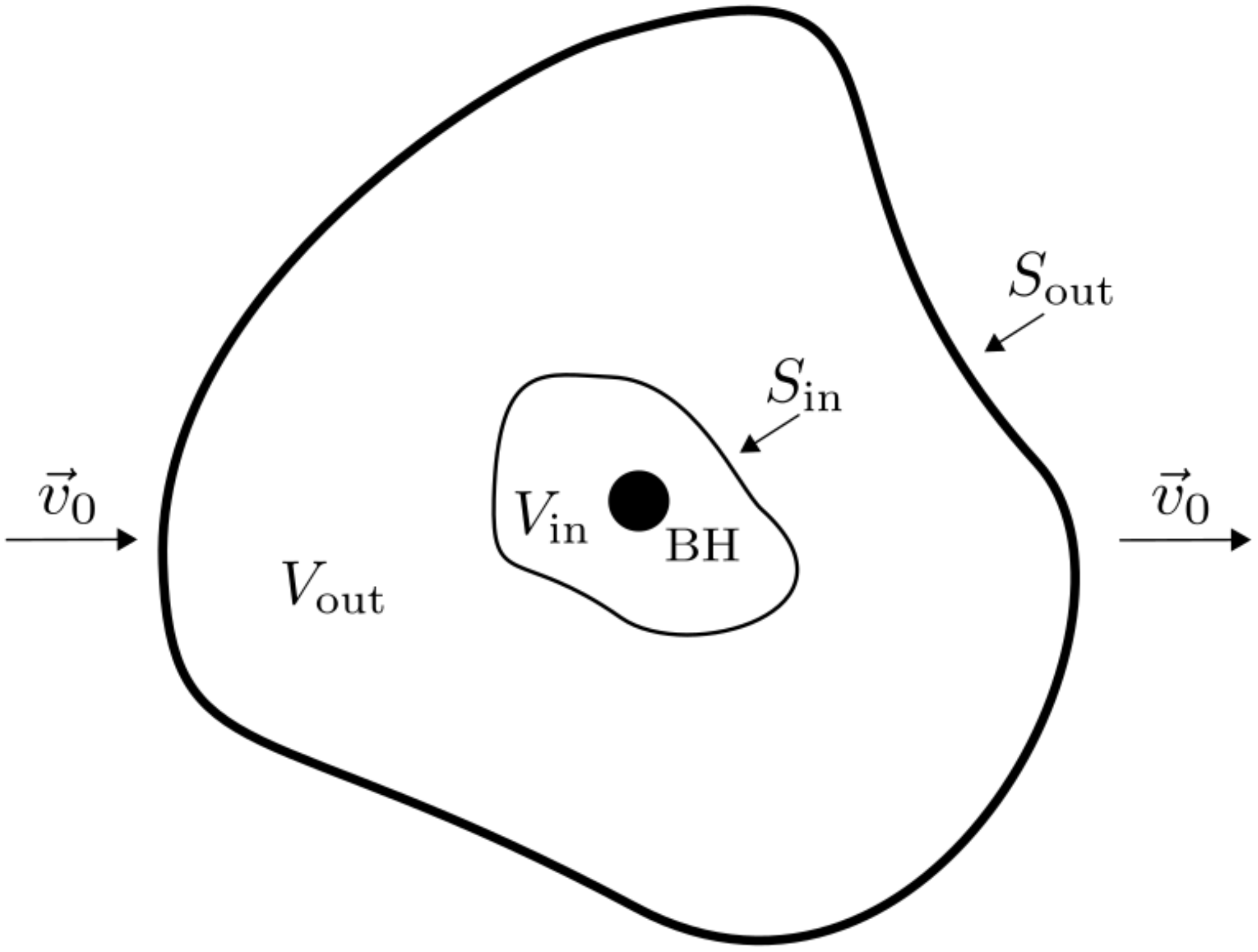}
\end{center}
\caption{
Inner and outer surfaces used in Eq.(\ref{eq:Fz-def}).
}
\label{fig_vol-friction}
\end{figure}

As the BH moves through the scalar-field cloud it is slowed down by a drag force,
often called dynamical friction.
By symmetry, this force $\vec F = F_z \vec e_z$ is directed along the $z$-axis.
As sketched in Fig.~\ref{fig_vol-friction},
let us consider an open subsystem formed by the BH and the scalar field inside a surface
$S_{\rm in}$ that encloses the BH, far enough from the horizon for Newtonian dynamics to hold
but close enough for its mass $M$ to be dominated by the BH mass $M_{\rm BH}$.
The surface $S_{\rm in}=\partial V_{\rm in}$  bounds a volume $V_{\rm in}$. Outside this volume the scalar cloud extends up to the soliton radius $R_{\rm sol}$, at a much greater distance.
This defines the outer volume $V_{\rm out}$.
Going back to physical coordinates, the change of momentum of this subsystem,
of volume $V_{\rm in}$, reads
\ba
\frac{d p_z}{dt} & = & {\cal G} M_{\rm BH} \int_{V_{\rm out}} d\vec r \rho(\vec r)
\frac{{\vec r} \cdot \vec e_z}{r^3} - \int_{\partial V_{\rm in}} \vec{dS} \cdot P \vec e_z \nonumber \\
&& - \int_{\partial V_{\rm in}} \vec{dS} \cdot \rho \vec v v_z .
\label{eq:Fz-def}
\ea
The first term, integrated over the volume $V_{\rm out}$  of the scalar cloud, is the usual dynamical
friction term due to the gravitational wake \citep{Mulder1983}, that is, the gravitational pull from
the scalar-field overdensity generated behind the BH through the deflection of the streamlines under
the BH gravity.
The second term, which is absent in collisionless media such as the stellar cloud considered
by Chandraskhar's classical study \cite{Chandrasekhar:1943ys},
is the pressure exerted by the outer cloud on the subsystem.
The third term is the contribution of the momentum flux through the surface $S_{\rm in}$. This last term is clearly related to the local influx of matter and therefore the infall of mass into the BH, i.e. accretion, but it vanishes if the flow is radial close to the BH.

In the limit of an infinite constant-density scalar cloud, the first gravitational term suffers
from the same divergence as the Newtonian gravitational force in an infinite homogeneous
universe, associated with the so-called ``Jeans swindle''. As usual, this can be cured by
integrating first over angles or by regularizing Newtonian gravity with a damping factor
$e^{-\kappa | \vec r - \vec r' |}$, taking the limit $\kappa\to 0$ at the end of the computations
\cite{Kiessling:1999eq}. This implies that a constant-density background does not contribute
and only the asymmetry of the perturbed scalar density field contributes, associated with the wake
behind the BH.

Considering the surface $S_{\rm in}$ as the inner boundary of the outer volume $V_{\rm out}$,
which changes the sign of $\vec{dS}$, and introducing the external surface $S_{\rm out}$ of the
scalar cloud itself, we can write the pressure term as
\ba
- \int_{\partial V_{\rm in}} \vec{dS} \cdot P \vec e_z & = &
\int_{\partial V_{\rm out}} \vec{dS} \cdot P \vec e_z
- \int_{S_{\rm out}} \vec{dS} \cdot P \vec e_z \nonumber \\
& = & \int_{V_{\rm out}} d\vec r \frac{\partial P}{\partial z}
- \int_{S_{\rm out}} \vec{dS} \cdot P \vec e_z ,
\ea
as $\partial V_{\rm out} = S_{\rm in} \cup S_{\rm out}$ and we used the divergence theorem
in the second line. If the scalar cloud is isolated in vacuum,
the pressure term vanishes at the cloud boundary. However, this is not the case if we choose
a surface $S_{\rm out}$ that is inside the scalar cloud, but large enough for the drag force
to have converged.
Then, noticing that the first term in Eq.(\ref{eq:Fz-def}) is also the opposite of the gravitational
attraction by the BH of the outer scalar cloud, we obtain
\ba
\frac{d p_z}{dt} & = & \int_{V_{\rm out}} d\vec r \left[ \rho \frac{\partial \Phi_{\rm BH}}{\partial z}
+ \frac{\partial P}{\partial z} \right] - \int_{S_{\rm out}} \vec{dS} \cdot P \vec e_z \nonumber \\
&& - \int_{\partial V_{\rm in}} \vec{dS} \cdot \rho \vec v v_z .
\label{eq:Fz-Phi_BH-P}
\ea
Far inside the soliton boundary, the steady-state Euler equation associated
with the continuity and Bernoulli equations (\ref{eq:continuity-Bernouilli}) reads
\be
\nabla \cdot ( \rho \vec v v_z ) = \rho \vec v \cdot \nabla v_z =
- \rho \frac{\partial\Phi_{\rm BH}}{\partial z} - \frac{\partial P}{\partial z} .
\ee
Substituting into Eq.(\ref{eq:Fz-Phi_BH-P}) we obtain
the drag force on the BH
\be
F_z = \frac{d p_z}{dt} = -  \int_{S_{\rm out}} \vec{dS} \cdot  \rho \vec v v_z
- \int_{S_{\rm out}} \vec{dS} \cdot P \vec e_z .
\label{eq:Fz-out}
\ee
This expression no longer depends on the inner surface $S_{\rm in}$, nor on the bulk of the volume
$V_{\rm out}$. Therefore, we can shrink the inner surface $S_{\rm in}$ towards the BH.
The first term expresses the conservation of momentum as for collisionless systems:
in the steady state, the momentum that enters the external boundary $S_{\rm out}$ is equal
to the gain of momentum of the BH (like in Eq.(\ref{eq:dot-M_BH-S}) the accretion of mass
by the BH is equal to the mass inflow through any enclosing surface $S$).
The second term takes into account the impact of the pressure, when the surface $S_{\rm out}$
is taken within the soliton cloud. The clear interpretation of Eq.(\ref{eq:Fz-out}) means that
it could have been used at once as the definition of the net drag force, in a steady state,
as in \cite{Lee:2011px} for the case of the isothermal gas.
The interest of the derivation above is to clarify its link with the expression (\ref{eq:Fz-def}),
which contains the more familiar gravitational wake term, associated with the usual meaning
of dynamical friction in the case of free particles.

\subsection{Relationship with the accretion rate}

As for the BH accretion rate, we can check that the dynamical friction converges to a finite
value that does not depend on the shape of the surface $S_{\rm out}$, in the large-distance limit.
Choosing for the surface $S_{\rm out}$ a distant sphere centered on the BH, as in the upper
panel of Fig.~\ref{fig_vol-accretion}, we obtain in dimensionless
variables the monopole contribution
$\hat F_z = - 4\pi \hat r^2 (\hat\rho v_r v_z + \cos\theta \hat P)_{\ell =0}$.
At large radius $r$, using the large-distance expansions derived from (\ref{eq:beta-large-r-a-b}),
we find that the factors $\hat r$ cancel out as expected and we obtain
$\hat F_z = v_0 \dot{\hat M}_{\rm BH}$, where $\dot{\hat M}_{\rm BH}$ is the BH accretion
rate obtained in (\ref{eq:dot-M_BH-sphere}).
Choosing instead for the surface $S_{\rm out}$ the elongated cylinder that follows the streamlines,
as for the computation (\ref{eq:dot-M-BH-cyl}) and as in the lower panel of
Fig.~\ref{fig_vol-accretion}, we find at once that the first term in
Eq.(\ref{eq:Fz-out}) gives $\hat F_z = \pi \hat\rho_0 v_0^2 (\hat b_-^2 - \hat b_+^2)
= v_0 \dot{\hat M}_{\rm BH}$.
An explicit computation from the expansions derived from (\ref{eq:beta-large-r-a-b})
shows that the pressure integral of the second term vanishes as $1/b$.
Therefore, we find that both computations give the same result,
\be
F_z = \dot M_{\rm BH} v_0 .
\label{eq:Fz-M_BH-v0}
\ee
Thus, the drag force is simply given by the product of the accretion rate and the relative
velocity.
Using Eq.(\ref{eq:MBdot-approx}) we obtain
\be
F_z \sim \rho_0 r_s^2 v_0/c_s^2 \sim {\cal G}^2 M_{\rm BH}^2 \rho_0 v_0 / c_s^2 .
\label{eq:Fz-approx}
\ee

We checked that our numerical computation of the scalar-field profile agrees with the prediction
(\ref{eq:Fz-M_BH-v0}). As explained in Sec.~\ref{sec:low-high-branches}, we match the
scalar-field cloud to the radial flow at the matching radius $r_{\rm m}$, somewhat beyond the
critical radius $r_c$ associated with the transition from the low-velocity to the high-velocity branch.
Choosing for the inner surface $S_{\rm in}$ the sphere of radius $r_{\rm m}$,
the second and third terms of Eq.(\ref{eq:Fz-def}) vanish by symmetry.
The first gravitational term reads
$(4\pi/3) {\cal G} M_{\rm BH} \int_{r_{\rm m}}^\infty dr \rho_{\ell = 1}$.
It is given by the dipole of the scalar-cloud density field (thus the unperturbed
background does not contribute), which decays as $1/r^2$ at large distance,
as seen in Eq.(\ref{eq:rho-even-odd-explicit}) and Fig.~\ref{fig_rho-ell},
and our numerical computation agrees with Eq.(\ref{eq:Fz-M_BH-v0}).

\subsection{Comparison with previous works and other systems}

\subsubsection{Accretion, gravitational wake and drag force}

There can be some confusion in the literature about the net drag force.
In the collisionless case, following Chandraskhar's classical work \cite{Chandrasekhar:1943ys},
it is usually called
dynamical friction and it is due to the long-range gravitational interaction between the perturber
(here the BH) and the distant stars of the stellar cloud. Summing over the changes of velocity
experienced during these distant encounters gives the well-known result
(\ref{eq:F-Chandrasekhar-all}) recalled below.
An alternative treatment is to compute the perturbation to the steady-state distribution of the stars
by the perturber \citep{Mulder1983}.
Indeed, by bending the trajectories of the distant stars the perturber generates
a gravitational wake behind it. The overdensity in this wake is then responsible for the deceleration
of the perturber \citep{Antonini2011}. Computing the gravitational pull from this overdense
wake gives back Chandraskhar's result.
This clearly corresponds to the first term in Eq.(\ref{eq:Fz-def}).
In such a system, there is no accretion nor pressure.
However, when there is accretion on the perturber (whether a BH or a massive star),
another source of momentum exchange is associated with the momentum deposited
by the accreted material, sometimes called a capture drag or accretion drag force.
In some studies, these two sources of momentum exchange are estimated separately,
found to be of the same order, typically given by the Chandraskhar expression
(\ref{eq:F-Chandrasekhar-all}) below, and one simply uses the latter formula
or adds both contributions.
However, naive estimates can create confusion and lead to double counting.
Indeed, our explicit computation (\ref{eq:Fz-M_BH-v0}) shows that the net drag force
is actually equal to a naive estimate of the momentum exchange associated with the accretion.
Then, one could wonder where the contribution associated with the long-range
gravitational interaction has disappeared.
Moreover, we find that the latter, given by the first term in Eq.(\ref{eq:Fz-def}),
is actually almost equal to (\ref{eq:Fz-M_BH-v0}).
The explanation is clearly seen in Eq.(\ref{eq:Fz-def}), where the first term is obviously the
dynamical friction associated with the long-range gravitational interaction, and the
third term the deposited momentum.
However, this latter quantity is much smaller than the naive estimate $\dot M_{\rm BH} v_0$
because matter falls almost radially onto the BH near the horizon, precisely because of the
gravitational interaction that bends the flow (combined with the fluid pressure).
Therefore, one cannot separate both effects.
Fortunately, Eq.(\ref{eq:Fz-out}) provides a simple expression for the net drag force
that does not need to separate between such a gravitational friction and a capture drag.
Therefore, we prefer to use the term net drag force to describe the total force felt by the BH,
which is the relevant quantity for practical purposes.

An advantage of the expression (\ref{eq:Fz-out}) is that it allows us to obtain the
analytical result (\ref{eq:Fz-M_BH-v0}), thanks to the relationship with the large-distance
expansions. This is quite useful as the accretion rates and especially
drag forces can be difficult to compute by numerical simulations, which can vary by factors
of a few or more depending on the numerical scheme  \cite{Petrich1989}.
Another key point is that it allows us to obtain the drag force as could be understood
from an effective theory point of view, thanks to the integral being performed at large
distance, while keeping account of all the nonlinear and relativistic effects close to the
Schwarzschild radius that actually determine the values of the accretion rate and of the dynamical
friction, as recalled in Sec.~\ref{sec:low-high-branches} and Eqs.(\ref{eq:rho-even-odd-explicit})
and (\ref{eq:B-F_star}).

\subsubsection{Classical systems}

As noticed in previous sections, the relation (\ref{eq:Fz-M_BH-v0}) explicitly shows
that the drag force vanishes with the accretion rate.
Indeed, in this case the potential flow is symmetric with respect to the $\hat z=0$ plane
and there can be no net force along the $\hat z$ axis.
Not surprisingly, in view of the hydrodynamical analogy derived in
Sec.~\ref{sec:isentropic} in the nonrelativistic regime, the drag force was also found to vanish
for the subsonic motion of a star in an isothermal gas, without accretion, from a linear
steady-state analysis \cite{Dokuchaev1964,Rephaeli1980,Ostriker:1998fa}.
However, our result (\ref{eq:Fz-M_BH-v0}) does not rely on a linear treatment.
It simply uses Gauss's theorem to write the total drag force in terms of the asymptotic
behavior of the fields at large distance (\ref{eq:Fz-out}). This merely expresses the conservation
of mass and momentum in a steady state.
Moreover, the value of the accretion rate itself (\ref{eq:dot-M-BH-radial}) involves a fully nonlinear
and relativistic treatment that extends from large radii down to the Schwarzschild radius
\cite{Brax:2019npi}.

When the accretion rate is nonzero the drag force no longer vanishes.
This is because the accretion onto the BH, associated with a radial inward velocity flow near
the BH horizon, breaks the symmetry with respect to the $\hat z=0$ plane.
This is clearly manifested by the turning point $r_{\rm turn}$ somewhat behind the BH,
associated with a local maximum of the dark matter density field in the wake.

The same relationship (\ref{eq:Fz-M_BH-v0}) between accretion and the drag force was
found in \citep{Lee:2011px} for a BH moving in an isothermal gas.
The proportionality to $\dot M_{\rm BH}$ is not surprising, as the dynamical friction
in a perfect fluid without accretion vanishes and the form of the relationship could
be expected from dimensional analysis.
However, the fact that the coefficient is unity is not obvious a priori (in the extreme case of
free particles, recalled in Eq.(\ref{eq:F-Chandrasekhar}) below, the drag force is actually nonzero
even when there is no accretion).
Despite this formal similarity with the case of the perfect gas, as noticed in
Eq.(\ref{eq:MBdot-approx}), the accretion rate for the scalar field is much smaller than for
the isothermal gas. This implies that the drag force is also much smaller.
Indeed, for the isothermal gas Ref.\cite{Lee:2011px} obtains
\be
\mbox{subsonic perfect gas:} \;\;\; F_{\rm perfect \; gas} \sim
{\cal G}^2 M_{\rm BH}^2 \rho_0 v_0/c_s^3 ,
\label{eq:Fz-perfect-gas}
\ee
and we find that both the accretion rate and the drag force are smaller
for the scalar dark matter by a factor $c_s/c \ll 1$, see our result (\ref{eq:Fz-approx}).
(The result (\ref{eq:Fz-perfect-gas}) was also obtained by \cite{Ostriker:1998fa}
using linear theory, without accretion but taking into account finite-time effects.
It is consistent with hydrodynamic simulations \citep{Kim2009}.)

The result obtained in Eq.(\ref{eq:Fz-M_BH-v0}) is also very different from the one obtained by
Chandrasekhar \cite{Chandrasekhar:1943ys} for free particles,
and confirmed by numerical simulations \citep{Binney1987,Antonini2011},
\be
\mbox{collisionless:} \;\;\;
F_{\rm free} \simeq 16\pi^2 C {\cal G}^2M_{\rm BH}^2\rho_0/v_0^2
\int_0^{v_0} dv \, v^2 f(v) ,
\label{eq:F-Chandrasekhar-all}
\ee
where the particle velocity distribution $f(\vec v)$ is normalized to unity
and assumed to be isotropic,
$C$ is the Coulomb logarithm $C \approx \ln(b_{\rm max}/{b_{\rm min}})$, with
$b_{\rm min} \sim {\cal G}M_{\rm BH}/v_0^2$ and $b_{\rm max}$
an infrared cut-off on the impact parameter $b_{\rm max}$, taken for instance
as the size of the cloud.
For a relative velocity $v_0$ that is smaller than the stellar cloud
velocity dispersion $c_s$, this gives
\be
v_0 < c_s : \;\;\;
F_{\rm free} \sim C {\cal G}^2M_{\rm BH}^2\rho_0 v_0/c_s^3
\label{eq:F-Chandrasekhar}
\ee
which shows the same scaling as the perfect-gas result (\ref{eq:Fz-perfect-gas}),
with the addition of the Coulomb logarithm, even though the relation with the
collisionless accretion rate (\ref{eq:accretion-free}) is very different,
$F_{\rm free} \sim \dot M_{\rm free} v_0 C c^2/c_s^2$.
Thus, even though the dynamical friction in the subsonic regime
is not zero, it is again much smaller than Chandrasekhar's result,
by a factor $c_s/(cC) \ll 1$.
Therefore, the scalar-field self-interactions are very efficient to reduce both the accretion rate and
the dynamical friction.

As explained in Sec.~\ref{sec:maps}, another impact of the scalar-field pressure,
which dominates over the BH gravity at large distance, is to screen the impact of the latter on the
dark matter distribution. As a result, the dark-matter density perturbation decays faster
at large distance, as $1/r^2$ for the odd component as seen in Eq.(\ref {eq:rho-even-odd-explicit}),
and there is no more large-radius divergence in a Coulomb logarithm factor.
Thus, the dynamical friction (\ref {eq:Fz-M_BH-v0}) does not depend on the size of the scalar
cloud.
There is no need for a small-scale cutoff either, because the odd components
of the product $r\rho_\ell$ decrease at small radii, as seen in
Fig~\ref{fig_rho-ell}.
The fact that our result (\ref{eq:Fz-M_BH-v0}) is finite and does not involve
small-scale nor large-scale cutoffs legitimates our steady-state computation.
Our predictions for the accretion rate and the dynamical friction should
apply as soon as sound waves have time to reach $r_\gamma$, so that transients
vanish and the steady state can be achieved.
This gives a time $t \sim r_\gamma / c_s \ll t_{\rm dyn}$, where we define
the typical dynamical timescale of the soliton $t_{\rm dyn} \sim R_{\rm sol}/c_s$,
as $R_{\rm sol} \gg r_\gamma$.
(Here we assume that the size of the soliton is much greater than the radius
$r_\gamma$ where the BH gravity and the soliton self-gravity are of the same order.)
Therefore, the steady-state predictions should be achieved on timescales much
shorter than the global dark-matter halo timescale.

\subsubsection{Fuzzy dark matter}

Another popular scalar-field scenario is the FDM model
\cite{Hu:2000ke}, where the self-interactions are negligible but the de Broglie wavelength is very large.
This corresponds to a scalar mass $m \sim 10^{-22}$ eV in order to have wavelike
effects up to galactic scales and possibly alleviate small-scale tensions of the CDM scenario.
Using the Coulomb scattering of a plane wave by an $1/r$ potential \citep{mott1965},
associated with the Schr\"odinger equation in the external Newtonian gravity of the BH,
Ref.\citep{Hui:2016ltb} finds that in this case the dynamical friction reads
\be
\mbox{FDM:} \;\;\; F_{\rm FDM} \sim \frac{{\cal G}^2M_{\rm BH}^2\rho_0}{v_0^2}
C(\beta, k R) ,
\label{eq:F-FDM-C}
\ee
with
\be
\beta = \frac{{\cal G} M_{\rm BH} m}{v_0} , \;\;\; k=m v_0 ,
\ee
$R$ is the size of the scalar cloud, and $C(\beta, k R)$ is given in terms of confluent
hypergeometric functions.
We write the size of the soliton as $R \sim 1/(m|\Phi_{\rm N}|^{1/2}) \sim 1/(m c_s)$,
where we define the velocity scale $c_s$ as $c_s^2 = |\Phi_{\rm N}|$ (this would
be the virial velocity in a classical system).
Then, the radius $r_{\rm sg}$ where the BH gravity is of the order of the soliton self-gravity
is $r_{\rm sg} \sim {\cal G} M_{\rm B}/c_s^2$.
This gives
\be
\beta \sim \frac{r_{\rm sg} c_s}{R v_0} , \;\;\;
k R \sim \frac{v_0}{c_s} .
\ee
Assuming the size of the soliton is much greater than the self-gravity radius $r_{\rm sg}$,
we typically have $\beta \ll 1$ and $kR \lesssim 1$ in the subsonic regime that we consider in this
paper. In this limit, Ref.\citep{Hui:2016ltb} gives $C \sim (kR)^2$ and we obtain
\be
\frac{r_{\rm sg}}{R} c_s \ll v_0 < c_s : \;\;\;
F_{\rm FDM} \sim \frac{{\cal G}^2M_{\rm BH}^2\rho_0}{c_s^2} ,
\label{eq:F-FDM}
\ee
see also \cite{Lancaster:2019mde,Annulli2020} for related studies.
This is greater than the classical result (\ref{eq:F-Chandrasekhar}) by a factor $c_s/v_0 > 1$.
This is because there is no integration over a distribution function $f(v)$ as in
Eq.(\ref{eq:F-Chandrasekhar-all}), with a cutoff at $v_0$ that expresses that higher-velocity
stars do not contribute to the dynamical friction. Indeed, as the BH moves inside the soliton,
which is a coherent state with a vanishing phase, all scalar-cloud velocities are zero
(as in a Bose-Einstein condensate) and below the BH velocity $v_0$.
However, there is still a milder cutoff $\propto (kR)^2 \propto (v_0/c_s)^2$, which gives the result
(\ref{eq:F-FDM}), related to the number of modes available in the spectrum of vibrations
of the scalar field.
The expression (\ref{eq:F-FDM}) agrees with the results of Ref.\cite{Berezhiani:2019pzd},
obtained from a different approach (their Eq.(4.12) for the case where the scalar cloud
radius is of the order of the Jeans length), and Ref.\cite{Hartman:2020fbg},
who used an hydrodynamical approach, which also include weak quartic self-interactions.
It also roughly agrees with numerical simulations \cite{Wang:2021udl}.
Thus, we find that scalar dark matter with self-interactions gives a dynamical friction
that is smaller than for FDM, by a factor $v_0/c \ll 1$.

If the size of the FDM cloud is much greater than the de Broglie wavelength,
$kR \gg 1$, one recovers the classical scaling (\ref{eq:F-Chandrasekhar}) \cite{Bar-Or:2018pxz}.
Indeed, for small de Broglie wavelength the scalar field behaves as a collection
of particles of size $\lambda_{\rm dB}$.

The result (\ref{eq:F-FDM-C}) was obtained from linear perturbation theory and for
Newtonian gravity \citep{Hui:2016ltb,Lancaster:2019mde}.
It should nevertheless remain a good approximation for FDM as long as the Compton
wavelength of the scalar cloud is much greater than the BH horizon, see also
\cite{Vicente:2022ivh}.
 In contrast, the large-mass limit studied in this paper assumes that $1/m \ll r_s$.
 Moreover, our result (\ref{eq:Fz-approx}), obtained for the motion
of a BH in the scalar cloud, involves fully nonlinear and relativistic effects close to the
Schwarzschild radius, which actually determine the accretion rate and the drag force.
We are thus investigating both different systems and different
regimes. In our case, the self-interactions dominate over the quantum pressure
and the Compton wavelength $1/m$ is much smaller than all astrophysical scales,
both the size of the scalar cloud and the BH horizon.

\section{Conclusion}
\label{sec:Conclusion}

Scalar-field dark matter scenarios where scalar self-interactions are not negligible differ from the fuzzy dark matter models as the induced self-interaction pressure can take over the role played by the quantum pressure for FDM. This leads to new equilibrium configurations, i.e. solitons, where the self-interaction pressure balances gravity and leads to a smooth core for dark matter halos. In addition, the self-interactions imply the existence of a non-vanishing speed of sound. This influences the dynamics of compact objects moving inside dark matter halos. In particular, this modifies the accretion rate and the dynamical friction of a moving BH inside a soliton.
Focusing on the subsonic regime, i.e. a relative velocity smaller than the
speed of sound, we have studied the dark matter flow for a system composed of a Schwarzschild BH and a SFDM soliton, in the case of a quartic self-interacting potential.

We have shown that in the large scalar mass limit, the dark matter behaves as
an isentropic potential flow, far from the Schwarzschild radius,
with a polytropic index $\gamma_{\rm ad}=2$.
This is a signature of the important collective effects associated with the
pressure generated by the self-interactions. As such, the system in the subsonic
regime is closer to the case of a perfect gas than to a collection of particles.

We find that in this low-velocity regime the flow remains close to a simple
linear approximation, where the velocity potential is described by a monopole
(set by the radial accretion at the Schwarzschild radius) and a dipole
(set by the uniform flow at large distance).
Going beyond this linear approximation, we solved the nonlinear equations of motion
with an iterative numerical scheme. We also derived large-distance
expansions up to subleading order. This allows us to obtain explicit results
for the accretion by the BH and its dynamical friction, as we show how they
are encoded in these large-distance expansions. However, our analysis includes both
nonlinear and relativistic effects, which play a key role close to the BH horizon.

Despite the similarity with the usual perfect fluid on large scales, the accretion
rate is much smaller than Bondi's result for a perfect gas, by a factor $c_s/c \ll 1$.
This is because the scalar field
departs from a perfect gas in the relativistic regime and the accretion rate
is set by the large-field regime, not far from the Schwarzschild radius,
where the self-interactions are able to significantly slow down the infall.
This leads in turns to a dynamical friction that is much smaller than
for the perfect gas, by the same factor $c_s/c$, as we recover the same relationship
between the accretion rate and the drag force, $F_z = \dot M_{\rm BH} v_0$.

As compared with the case of free collisionless particles, we obtain an accretion rate that is
much greater, by a factor $c/c_s \gg 1$. However, the dynamical friction is still much smaller
than Chandrasekhar's result, by a factor $c_s/(cC) \ll 1$, where $C$ is the Coulomb logarithm,

We also find that the dynamical friction for self-interacting dark matter, in this large
scalar-mass regime, is smaller than for FDM, by a factor $v_0/c \ll 1$.

The smaller dynamical friction associated with such self-interacting scalar dark matter
might result in a decrease of the dephasing of the emission frequency of gravitational waves as compared to FDM and CDM. Also, SFDM could play a role in the Fornax globular cluster timing problem if dynamical friction is reduced  \cite{Bar:2021jff}.
However, the computation presented in this paper only applies to the case of a BH, with its
specific boundary condition at the Schwarzschild radius. For globular clusters and stellar objects,
with negligible accretion, we recover zero dynamical friction at leading order,
as found in the subsonic regime for perfect fluids in \cite{Ostriker:1998fa}, because of the
hydrodynamic analogy obtained in the nonrelativistic regime in Sec.~\ref{sec:isentropic}.
Nevertheless, a more accurate treatment, taking into account the perturbation to the
fluid self-gravity, is expected to show a small but nonzero dynamical friction as found
by \cite{Berezhiani:2019pzd}. This is left for future studies.
Other extensions of this work would be to consider other self-interactions,
beyond the quartic case, or scalar field models with non-standard kinetic
potentials, and to generalize the analysis to the case of a rotating BH.

\section{Acknowledgements}
Ph.B. acknowledges support from the European Union’s Horizon 2020 research and innovation programme under the Marie Skłodowska -Curie grant agreement No 860881-HIDDeN.

\appendix

\section{Green functions}
\label{sec:Green}

In the appendices that follow we drop the hats for simplicity of notation.

To go beyond the linear flow presented in Sec.~\ref{sec:linear-flow}, we split
the nonlinear equation (\ref{eq:flow-k_+-hatbeta}) as the system of two equations
(\ref{eq:system-source}) and we solve this system with an iterative scheme.
The source $S$ is readily obtained from the flow $\beta$ with the second equation.
The flow $\beta$ is obtained from the source $S$ by solving the first equation with the
help of the appropriate Green function $G(\vec r,\vec r^{\;'})$, which satisfies
$\nabla \cdot ( k_+^2 \nabla G ) = \delta_D ( \vec r - \vec r^{\;'})$.
Thus, the first equation in (\ref{eq:system-source}) gives
$\beta = \beta^L + \int d \vec r^{\;'} \; G(\vec r,\vec r^{\;'}) S(\vec r^{\;'})$,
where $\beta^L$ is the linear flow (\ref{eq:beta_L-def}).
Expanding the Green function in  spherical harmonics,
\be
G(\vec r,\vec r^{\;'}) = \sum_{\ell , m} G_\ell(r,r') Y^m_\ell(\theta',\varphi')^*
Y^m_\ell(\theta,\varphi) ,
\ee
while we decompose the phase $\beta$ and the source $S$ in  Legendre polynomials, as in
\be
\beta(r,\theta) = \sum_\ell \beta_\ell(r) P_\ell(\cos\theta) ,
\ee
we obtain
\be
\beta_\ell = \beta^L_\ell + \int_{r_{\rm m}}^\infty dr' \; r'^2 G_\ell(r,r') S_\ell(r') .
\label{eq:beta-Green}
\ee
As $\beta^L$ already matches the boundary conditions, the Green function must become
negligible at $r_{\rm m}$ and at large radii.
Thus, we take
\ba
&& \hspace{-0.3cm} r < r' : \;\;\; G_\ell(r,r') = w_\ell \, G_\ell^+(r) \, G_\ell^-(r') ,
\nonumber \\
&& \hspace{-0.3cm} r > r' : \;\;\; G_\ell(r,r') = w_\ell \, G_\ell^-(r) \, G_\ell^+(r') ,
\ea
where
\be
w_\ell = \frac{3}{2 (r+\gamma r^2) [ G_\ell^+(r) G_\ell^{-'}(r) - G_\ell^{+'}(r) G_\ell^-(r) ] }
\nonumber
\ee
is a constant thanks to the Wronskian theorem.
At the inner boundary $r_{\rm m}$, this gives $\frac{\partial G_0}{\partial r}(r_{\rm m}) = 0$
and we recover the radial velocity.
For the modes $\ell \neq 0$, the radial and angular velocities are not exactly zero at
$r_{\rm m}>0$. This was also the case for the linear dipole (\ref{eq:beta1-2F1}),
and we only require that they are negligible.
This is because we cut our numerical solution at the radius $r_{\rm m}$, somewhat above
the Schwarzschild radius, where the flow is already dominated by the radial accretion
but angular velocities are not exactly zero.
We can check in Fig.~\ref{fig_dbeta_dr} that this is a good approximation.

\section{Mode coupling at large radii}
\label{sec:mode-coupling-large-radii}

Here we present an alternative approach to that of Sec.~\ref{sec:Mach-large-radii} to show
how the nonlinear mode couplings, associated with the cubic nonlinearity
in Eq.(\ref{eq:flow-k_+-hatbeta}), generate the large-distance odd and even tails
(\ref{eq:beta-large-r-a-b}) for all Legendre multipoles, starting from the linear seed
(\ref{eq:beta_L-def}).
This follows the spirit of the iterative numerical scheme applied to the system
(\ref{eq:system-source}).

\subsection{Odd multipoles}
\label{sec:Odd-multipoles}

\subsubsection{Constant velocity-potential tail for odd multipoles}

The cubic nonlinearity in $\beta$ in Eq.(\ref{eq:flow-k_+-hatbeta}) generates some mode
coupling between Legendre multipoles.
We can estimate this effect at large radii, where the velocity is almost equal to
$\vec v_0$ and $\beta$ is dominated by the dipole in expression (\ref{eq:beta_L-def}).
First, we note that the linear solution (\ref{eq:beta_L-def}) behaves at large radii as
Eq.(\ref{eq:2F1-expansion}).
At leading order for $r\to\infty$, the source $S$ in (\ref{eq:system-source}) is dominated
by the contribution from two powers of the leading term $v_0 r$ and one power of the
subleading term $-v_0/2\gamma$ from $\beta_1^L$.
This gives a contribution of order $1/r^2$ to $S$. Terms of order $1/r^2$ are also generated
in the left-hand side in the first eq.(\ref{eq:system-source}) by constant terms in $\beta$
and by the contribution from the leading $v_0 r$ term paired with the contribution $2/(3r)$
to the kernel $k_+^2$ in (\ref{eq:gamma-def}).
This means that mode couplings generate a constant tail for all odd multipoles,
as repeated action of the cubic nonlinearity spreads power from the dipole to all higher-order
odd multipoles.
Thus, we write for the odd part of the phase $\beta_{\rm odd}$ a large-distance expansion
of the form (\ref{eq:beta-large-odd-even})-(\ref{eq:beta-large-r-a-b}).
For the linear flow, we have $a_1=-v_0/(2\gamma)$ and all other multipoles are zero.
Collecting all the terms of order $1/r^2$ in Eq.(\ref{eq:flow-k_+-hatbeta}), we obtain
\ba
&& \hspace{-0.5cm} - \frac{\gamma}{r^2} \sum_{n \; {\rm odd}} a_n n(n+1) P_n - \frac{v_0 \cos\theta}{r^2}
= \frac{2 v_0^2}{r^2} \sum_{n \; {\rm odd}} a_n n (n+1)
\nonumber \\
&& \hspace{-0.3cm} \times \left[ \frac{(n+2)^2}{(2n+1) (2n+3)} P_{n+2}
+ \frac{(n-1)^2}{(2n-1) (2n+1)} P_{n-2} \right. \nonumber \\
&& \hspace{-0.3cm} \left. - \frac{2n^2+2n-1}{(2n-1) (2n+3)} P_n \right]
- \frac{v_0^2}{r^2} \sum_{n \; {\rm odd}} a_n n(n+1) P_n .
\ea
For the linear flow, which neglects the right-hand side, we recover $a_1=-v_0/(2\gamma)$.
Collecting the coefficient of each Legendre polynomial $P_n$, this gives the recursion
for odd integers $n \geq 1$,
\ba
&& - 3 k_0^2 a_n - v_0 \delta_{n,1} = 4 v_0^2
\left[ \frac{n (n-2) (n-1)}{(n+1) (2n-3) (2n-1)} a_{n-2} \right.
\nonumber \\
&&  \left. + \frac{(n+1) (n+2) (n+3)}{n (2n+3) (2n+5)} a_{n+2}
- \frac{2n^2+2n-1}{(2n-1) (2n+3)} a_n \right] .
\nonumber \\
&& \label{eq:a_ell_odd_recursion}
\ea
Thus, because of the mode couplings induced by the cubic nonlinearity,
a nonzero $a_1$ generates nonzero values for all odd multipoles.

Defining the parameter
\be
\xi = \frac{3 k_0^2}{v_0^2} = 4 \frac{c_s^2}{v_0^2} ,
\label{eq:xi-def}
\ee
where $c_s$ was defined in Eq.(\ref{eq:vs-def}),
this recursion becomes at large $n$
\be
n \gg 1 : \;\;\; - \xi a_n = a_{n-2} + a_{n+2} - 2 a_n .
\label{eq:recursion-large-ell}
\ee
This has two independent solutions of the form
\be
a_n = y_+^{n/2} \;\;\; \mbox{and} \;\;\; a_n = y_-^{n/2} ,
\label{eq:a_ell_y+_y-}
\ee
with
\ba
&& 0 < \xi < 4 : \;\;\; y_\pm = \frac{-(\xi-2)\pm i \sqrt{4-(\xi-2)^2}}{2} ,  \nonumber \\
&& \mbox{hence} \;\;\; y_\pm = e^{\pm i \zeta} \;\; \mbox{with} \;\;
\zeta = \arccos(1 - \xi/2) ,
\label{eq:y-pm-high-v0}
\ea
and
\be
\xi > 4 : \;\;\; y_\pm = \frac{-(\xi-2)\pm \sqrt{(\xi-2)^2-4}}{2} .
\label{eq:y-pm-low-v0}
\ee

\subsubsection{Fast decaying multipole amplitudes for \texorpdfstring{$v_0 < c_s$}{Lg}}
\label{sec:low-velocity-odd}

For small velocities, $\xi > 4$, that is $v_0 < c_s$,
we have $|y_-| > 1 > |y_+|$ and the requirement to have
a well-defined multipole expansion selects the decaying solution $y_+^{n/2}$.
The recursion (\ref{eq:a_ell_odd_recursion}), starting from the second equation at
$n = 3$,
defines all $a_n$ with $n \geq 5$, with $a_1$ and $a_3$ being still undetermined,
as this is a linear difference equation of second order (three-terms recursion),
with two independent solutions. The selection of the decaying solution $y_+^{n/2}$
then selects the ratio $a_3/a_1$. Substituting $a_3$ in terms of $a_1$ in the first equation
(\ref{eq:a_ell_odd_recursion}), with $n=1$, then determines $a_1$.
Thus, all coefficients $a_n$ are uniquely determined.
For $v_0 \to 0$, the amplitude of the coefficients $a_n$ shows a fast decrease with
$n$, as $|a_n | \sim |y_+|^{n/2} \sim v_0^n$, and the right-hand side in the
first equation (\ref{eq:a_ell_odd_recursion}) becomes negligible. Thus, we recover as
expected the linear flow (\ref{eq:2F1-expansion}),
\be
v_0 \ll c_s : \;\;\; a_1 \simeq - \frac{v_0}{3 k_0^2} \simeq - \frac{v_0}{2\gamma} , \;\;\;
a_n \sim v_0^{n} .
\label{eq:a-ell-v0-small}
\ee
These results agree with the explicit expression
(\ref{eq:deltabeta-odd-large-r-low-v0}).

Thus, whereas for the linear flow (\ref{eq:beta_L-def}) multipoles beyond the dipole are exactly
zero, the mode couplings lead to a constant value for these odd multipoles at large $r$,
hence multipoles of the angular velocity that only decay as $1/r$, and not with a power of
$1/r$ that grows with $n$.
This is quite different from the large-distance behavior of the linear modes $G_n^-$
in (\ref{eq:G-large-r}) of the operator $\nabla [ k_+^2 \nabla ( \cdot ) ]$.
This is because the source $S$ is not strongly peaked at a given scale, such as $r_\gamma$,
so that the integral over the Green function in (\ref{eq:beta-Green}) is not peaked around
a finite range of $r'$. Instead, significant contributions arise up to $r' \sim r$.
This means that the results (\ref{eq:deltabeta-odd-large-r-low-v0}) and (\ref{eq:a_ell_odd_recursion})
are robust and the large-distance behavior of the phase $\beta$ is not sensitive to its
behavior at small radii.

\subsubsection{Appearance of a singularity for \texorpdfstring{$v_0 > c_s$}{Lg}}
\label{sec:singularity-v0>vs}

In agreement with the change of flow property found in (\ref{eq:subsonic})-(\ref{eq:supersonic}),
from the subsonic to the supersonic regime, we briefly note that the threshold $c_s$ can be
recovered from the result (\ref{eq:a_ell_odd_recursion}).
For $v_0 > c_s$, we have at large $n$ the two independent real solutions
${\rm Re}(y_+^{n/2}) = \cos(n \zeta/2)$ and
${\rm Im}(y_+^{n/2}) = \sin(n \zeta/2)$.
At large $n$ the Legendre polynomials take the asymptotic form
\be
n \to \infty : \;\;\; P_n(\cos\theta) = \sqrt{\frac{2}{\pi n \sin\theta}}
\cos\left[ \left(n+\frac{1}{2}\right) \theta - \frac{\pi}{4} \right]
\label{eq:Legendre-large-ell}
\ee
for $0 < \theta < \pi$, and both solutions $\cos(n\zeta/2)$ and $\sin(n\zeta/2)$ lead to
divergent series in $n$ when $\theta=\zeta/2$.
This agrees with the singularity of Eq.(\ref{eq:deltabeta-odd-large-r-low-v0}) at the threshold
$c_s$, that is, at $\mu = 0$.
Thus, we recover that for $v_0 > c_s$ a shock appears and the flow is no longer
a smooth perturbation around the linear flow (\ref{eq:beta_L-def}).
We leave the study of this regime for a companion paper.

\subsection{Even multipoles}
\label{sec:Even-multipoles}

\subsubsection{\texorpdfstring{$1/r$}{Lg} velocity-potential tail for even multipoles}

As seen in the previous appendix \ref{sec:Odd-multipoles}, at large radii the
nonlinear mode couplings generate contributions to all odd multipoles as soon as the
dipole is nonzero.
This is because there is a partial decoupling between odd and even multipoles in the
nonlinear system (\ref{eq:system-source}). The linear operator
$\nabla [ k_+^2 \nabla \beta ]$ keeps the parity of $\beta$ whereas the cubic
source term $\nabla \cdot [ (\nabla\beta)^2 \nabla\beta ]$ requires one or three
odd multipoles to generate an odd multipole, and one or three even multipoles to generate
an even multipole. In other words, thinking in terms of an iterative scheme,
starting with a phase $\beta$ that has no even component, the cubic source term
will never generate even multipoles. Then, a fixed point reached by iterating the system
(\ref{eq:system-source}) (assuming the iteration converges) will only contain odd
multipoles.
If there is an initial seed for even multipoles, then all even order mutlipoles will
be generated by the cubic nonlinearity, and they will mix with the odd multipoles
through products of two odd terms with one even term.
However, because of the partial decoupling, the final amplitude of the even terms
will be proportional to the initial seed (similarly, we can check in
Eq.(\ref{eq:a_ell_odd_recursion}) that all odd coefficients $a_n$ vanish if
$v_0=0$).

Thus, because the linear flow (\ref{eq:beta0-L}) has an even component $\beta_0$
that decays as $1/r$ at large radii, we write for the even part of the phase $\beta_{\rm even}$
a large-distance expansion of the form (\ref{eq:beta-large-odd-even})-(\ref{eq:beta-large-r-a-b}).
The even components decay faster than the odd components, which go to a constant
(except for the leading growing dipole term), because of the partial decoupling explained above.
For the linear flow, we have $b_0 \sim 1/\gamma$ and all other multipoles are zero.
Collecting all the terms of order $1/r^3$ in Eq.(\ref{eq:flow-k_+-hatbeta}), we obtain
\ba
&& \hspace{-0.4cm} - \frac{\gamma}{r^3} \sum_{n \; {\rm even}} b_n n(n+1) P_n
= \frac{2 v_0^2}{r^3} \sum_{n \; {\rm even}} b_n \nonumber \\
&& \hspace{-0.3cm} \times \biggl[
\frac{(n+1)^2 (n+2) (n+3)}{(2n+1) (2n+3)} P_{n+2}
+ \frac{n^2 (n-2) (n-1)}{(2n-1) (2n+1)} P_{n-2} \nonumber \\
&& \hspace{-0.3cm} - \frac{n (n+1) (2n^2+2n-1)}{(2n-1) (2n+3)} P_n \biggl]
- \frac{v_0^2}{r^3} \sum_{n \; {\rm even}} b_n n(n+1) P_n . \nonumber \\
&&
\label{eq:b_ell_even_sum}
\ea
For the linear flow, which neglects the right-hand side, we recover that $b_n=0$
for $n \geq 2$ while $b_0$ can take any value (set by the small-radius boundary
condition).
Collecting the coefficient of each Legendre polynomial $P_n$, this gives the recursion
for even integers $n \geq 2$,
\ba
&& - 3 k_0^2 b_n = 4 v_0^2
\left[ \frac{(n-1)^2}{(2n-3) (2n-1)} b_{n-2} \right.
\nonumber \\
&&  \left. + \frac{(n+2)^2}{(2n+3) (2n+5)} b_{n+2}
- \frac{2n^2+2n-1}{(2n-1) (2n+3)} b_n \right] .
\nonumber \\
&& \label{eq:b_ell_even_recursion}
\ea
The equation obtained from (\ref{eq:b_ell_even_sum}) at $n=0$ is automatically
satisfied because all terms include the prefactor $n(n+1)$, which cancelled
out in (\ref{eq:b_ell_even_recursion}) for $n \geq 2$.
Again, a nonzero $b_0$ generates nonzero values for all even multipoles.
At large $n$ we recover the same recursion as in (\ref{eq:recursion-large-ell}),
with again the two independent solutions (\ref{eq:a_ell_y+_y-}).

\subsubsection{Fast decaying multipole amplitudes for \texorpdfstring{$v_0 < c_s$}{Lg}}

At low velocities, $v_0 < c_s$, we thus recover the decaying solution
$y_+^{n/2}$ from Eq.(\ref{eq:y-pm-low-v0}). However, because the relation
(\ref{eq:b_ell_even_recursion}) at $n=0$ was automatically satisfied as $0=0$,
the series of even multipoles is not uniquely determined and depends on the unconstrained
monopole coefficient $b_0$, which is set by the matching to the behavior at small radii.
For $v_0 \to 0$ we recover the linear flow (\ref{eq:beta0-L}) as
\be
v_0 \ll c_s : \;\;\; b_n \sim v_0^{n} / \gamma .
\label{eq:b-ell-v0-small}
\ee
Again, this $1/r$ tail is not sensitive to the behavior of the phase $\beta$ at
small radii, except for its overall normalization.

At large velocities, $v_0 > c_s$, we recover the oscillating modes
(\ref{eq:y-pm-high-v0}) and the singularity analysed in App.~\ref{sec:singularity-v0>vs}.
This agrees with the singularity at $\mu \to 0$ of the explicit expression
(\ref{eq:deltabeta-even-large-r}) and signals the appearance of a shock and the fact that
the flow is no longer a smooth perturbation of the linear flow.

\bibliography{friction_BH}

\end{document}